\documentclass[11pt,a4paper]{article}
\usepackage{jcappub}
\usepackage{latexsym,hyperref}
\usepackage{jcappub}
\input{epsf}
\newcommand{\beq}{\begin{equation}}
\newcommand{\eeq}{\end{equation}}
\newcommand{\bea}{\begin{eqnarray}}
\newcommand{\eea}{\end{eqnarray}}
\newcommand{\bi}{\begin{itemize}}
\newcommand{\ei}{\end{itemize}}
\newcommand{\bfi}{\begin{figure}[!t]
\epsfxsize=7cm
\epsffile}
\newcommand{\bfib}{\begin{figure}[htb]
\epsfxsize=9cm
\epsffile}
\newcommand{\bfig}{\begin{figure*}[htb]
\epsfxsize=12cm
\epsffile}
\newcommand{\efi}{\end{figure}}
\newcommand{\efib}{\end{figure}}
\newcommand{\efig}{\end{figure*}}

\newcommand{\ga}{\gtrsim}

\newcommand{\apjl}{ApJL}
\newcommand{\mnras}{MNRAS}
\newcommand{\jcap}{JCAP}
\newcommand{\prd}{Physical Review D}

\newcommand{\apj}{APJ}
\newcommand{\nat}{Nature}
\newcommand{\bfs}{\mbox{\boldmath$s$}}
\newcommand{\bfr}{\mbox{\boldmath$r$}}
\newcommand{\bfx}{\mbox{\boldmath$x$}}
\newcommand{\bfk}{\mbox{\boldmath$k$}}
\newcommand{\bfv}{\mbox{\boldmath$v$}}
\newcommand{\bfu}{\mbox{\boldmath$u$}}

\newcommand{\hompc}{\,h\,{\rm Mpc}^{-1}}

\newcommand{\ompc}{\,{\rm Mpc}^{-1}}

\def\be{\begin{equation}}
\def\ee{\end{equation}}
\def\ba{\begin{eqnarray}}
\def\ea{\end{eqnarray}}

\def\nn{\nonumber}

\title{Hybrid modeling of redshift space distortions}
\author[a]{Yong-Seon Song}
\author[b,a]{, Yi Zheng}
\author[c,d]{, Atsushi Taruya}
\author[a,e]{and Minji Oh}

\affiliation[a]{Korea Astronomy and Space Science Institute, 776, Daedeokdae-ro, Yuseong-gu, Daejeon 34055, Republic of Korea}
\affiliation[b]{School of Physics, Korea Institute for Advanced Study, Hoegiro 85, Seoul 02455, Korea}
\affiliation[c]{Center for Gravitational Physics, Yukawa Institute for Theoretical Physics, Kyoto University, Kyoto 606-8502, Japan}
\affiliation[d]{Kavli Institute for the Physics and Mathematics of the Universe (WPI), The University of Tokyo Institutes for Advanced Study, The University of Tokyo, 5-1-5 Kashiwanoha, Kashiwa, Chiba 277-8583, Japan}
\affiliation[e]{University of Science and Technology, Daejeon 34113, Korea}

\emailAdd{ysong@kasi.re.kr}
\emailAdd{yizheng@kasi.re.kr}
\emailAdd{ataruya@yukawa.kyoto-u.ac.jp}
\emailAdd{minjioh@kasi.re.kr}

\abstract{The observed power spectrum in redshift space appears distorted due to the peculiar motion of galaxies, known as redshift-space distortions (RSD). While all the effects in RSD are accounted for by the simple mapping formula from real to redshift spaces, accurately modeling redshift-space power spectrum is rather difficult due to the non-perturbative properties of the mapping. Still, however, a perturbative treatment may be applied to the power spectrum at large-scales, and on top of a careful modeling of the Finger-of-God effect caused by the small-scale random motion, the redshift-space power spectrum can be expressed as a series of expansion which contains the higher-order correlations of density and velocity fields. In our previous work [JCAP 8 (Aug., 2016) 050], we provide a perturbation-theory inspired model for power spectrum in which the higher-order correlations are evaluated directly from the cosmological $N$-body simulations. Adopting a simple Gaussian ansatz for Finger-of-God effect, the model is shown to quantitatively describe the simulation results. Here, we further push this approach, and present an accurate power spectrum template which can be used to estimate the growth of structure as a key to probe gravity on cosmological scales. Based on the simulations, we first calibrate the uncertainties and systematics in the pertrubation theory calculation in a fiducial cosmological model. Then, using the scaling relations, the calibrated power spectrum template is applied to a different cosmological model. We demonstrate that with our new template, the best-fitted growth functions are shown to reproduce the fiducial values in a good accuracy of 1 \% at $k<0.18 \hompc$ for cosmologies with different Hubble parameters.}

\begin{document}
\maketitle
\flushbottom

\section{Introduction}
\label{sec:Intro}

The large-scale structure observed via spectroscopic measurements exhibits anisotropies along the line-of-sight direction. This is caused by the peculiar velocity of galaxies, and is referred to as the redshift-space distortions (RSD) ~\cite{Jackson72,Sargent77,Peebles80,Kaiser87,Peacock94,Ballinger96}. While the RSD complicates the cosmological interpretation of the galaxy clustering data, on large scales, the size of the anisotropies is known to be sensitive to the growth of structure \cite{Kaiser87}, and can be used to test gravity on cosmological scales (e.g., \cite{Linder05,Guzzo08,Percival09,Song09,Beutler_bias,BOSSRD2012b,Song15a,Song15b}). This is one of the main reasons why there are various projects aiming at precisely measuring RSD which will uncover a large cosmic volume.

Future measurement of RSD is expected to further improve the statistical precision, with which we will be able to tightly constrain or possibly detect the modification to gravity on large scales. Toward a high-precision test of gravity, however, the theoretical template of the redshift-space power spectrum or correlation function,  as basic quantities to measure the growth functions, also needs to be improved. While the effects of RSD are solely accounted for by the simple mapping formula from the real to redshift spaces,  due to the non-perturbative nature of the mapping, the applicable range of linear theory prediction is fairly limited, and corrections coming from the gravitational clustering and RSD become rather significant even at $k\lesssim0.1\,h$Mpc$^{-1}$. In particular, the so-called Finger-of-God (FoG) effect, arising from the random motion of galaxies associated with small-scale clustering, appears non-perturbative and it leads to a strong damping behavior in the large-scale amplitude of power spectrum ~\cite{Fisher95,Heavens98,White01,Seljak01,Kang02,Tinker06,Tinker07,Scoccimarro04,Matsubara08a,Matsubara08b,Desjacques10,Taruya10,Taruya13,Matsubara11,Okumura11a,Okumura11b,Sato11,Jennings11b,Reid11,Seljak11,Kwan12,Zhangrsd,Zheng13,Ishikawa14,White15,Okumura15b,Jennings16,Bianchi15,Bianchi16,Simpson16,HandRSD17}.

Nevertheless, there is a way to perturbatively describe the theoretical template of redshift-space power spectrum, while keeping the non-perturbative FoG effect under control. Among various approaches or proposals for redshift-space power spectrum, in this paper, we shall consider the approach by \cite{Taruya10,Taruya13}. A crucial point of this approach is to decompose the contributions into non-perturbative part and the terms which can be evaluated with perturbation theory (PT) calculation, starting with the exact expression. Based on the simple proposition, the non-perturbative damping term is then separated out from the rest of the contributions, for which we can apply the PT calculation. As a result, on top of the factorized FoG damping term, the redshift-space power spectrum can be expressed as an infinite series of correction terms which contains the higher-order correlations of density and velocity fields. Ref.~\cite{Taruya10} derived the first non-trivial PT corrections relevant at next-to-leading order (i.e., one-loop). Later, Ref.~\cite{Taruya13} has extended to include the correction terms relevant for next-to-next-to-leading order calculations (i.e., two-loop).

On the other hand, Ref.~\cite{Zheng16a} has investigated the validity and consistency of this approach by directly evaluating each of the correction terms with $N$-body simulations. While the approach proposed by Refs.~\cite{Taruya10,Taruya13} is shown to be a good description, they found an important higher-order correction, for which the PT calculations by Ref.~\cite{Taruya13} was unable to quantitatively describe. Adding the numerically calibrated correction terms, the model of RSD is shown to reproduce the measured redshift-space power spectrum quite well.

In this paper, we further push this treatment toward a more accurate theoretical template of redshift-space power spectrum. A goal of this paper is to present the template which enables us to measure the growth functions in 1\% accuracy (we will later define the growth functions for density and velocity fields, which we denote by $G_\delta$ and $G_\Theta$, respectively). For this purpose, any small flaw in the model prediction has to be eliminated. This is also the case for PT calculations. As increasing the wavenumber, a small but non-negligible higher-order correction become important, and a more elaborate PT calculation is required for a quantitative estimate of such a contribution. Further, as it has been recently advocated, the frequently used single-stream approximation for the PT calculation is rather sensitive to the small-scale modes, and a proper way to control the UV-sensitive behavior is crucial. Hence, in this paper, we adopt the hybrid treatment combining the PT calculation and N-body simulations. That is, we use simulations to calibrate the systematics in the PT calculations, and to measure the correction terms of the redshift-space power spectrum. While we shall perform these calibration and measurements in a fiducial cosmological model, making use of the scaling relation, the numerically tabulated theoretical template can be also applied to different cosmological models. We demonstrate that with our template, the best-fitted growth functions, $G_\delta$ and $G_\Theta$, are shown to reproduce the fiducial values remarkably well in 1\% accuracy at $k < 0.18 \hompc$ for cosmologies with different Hubble parameters.

This paper is organized as follows. In Sec.~2, we begin by briefly describing the PT-inspired model of redshift-space power spectrum based on Refs.~\cite{Taruya10,Taruya13,Zheng16a}. Then, we discuss how to calibrate the systematics in the PT calculation and to measure the higher-order corrections from $N$-body simulations. The measured or calibrated results from the simulations are combined with PT predictions, and using a simple scaling relation, we show that the tabulated template in fiducial cosmology can be used to predict the power spectrum in different cosmological model. In Sec.~3, our newly constructed template is tested and the accuracy of the growth rate measurement is checked. Finally, Sec.~4 is devoted to conclusion and discussion.

\section{Hybrid modeling of redshift-space power spectrum}
\label{sec:estimation}

In testing gravity with RSD, we are mostly interested in a possible deviation from the standard $\Lambda$CDM model which appears manifest at late-time evolution of the Universe. In such a situation, the Universe basically follows the standard scenario of both the cosmic expansion and structure formation at least until the time of last scattering surface. Then, the broadband shape of the matter power spectrum, including the acoustic signature originated from the sound wave of the primeval baryon-photon fluid system, is basically the same one as in the $\Lambda$CDM model, and is determined by the cosmic microwave background experiments. Late-time evolution of the power spectrum may differ from each other among different dark energy models, but the broad-band shape remains unchanged as long as we consider the linear stage of structure formation, and this could also hold for a class of modified gravity models. In what follows, we adopt the power spectrum determined by the Planck $\Lambda$CDM model as our fiducial model, given by the parameters: $n_S=0.97\pm 0.0060$, $\Omega_{\rm b}h^2=0.022\pm0.00023$, $\Omega_{\rm c}h^2=0.12\pm 0.0022$, $h=0.67$ and $A_S^2=2.3\times 10^{-9}$~\cite{PLANK2015}.  In addition, assuming the flat cosmology, we also examine four other cosmological models with different value of $h$: $h=(0.57, 0.62,0.72,0.77)$, fixing $\omega_b$ and $\omega_c$.

\subsection{Perturbative treatment of redshift-space power spectrum and beyond}

As we mentioned in Sec.~1, all the effects in RSD is accounted for by the simple relation between real and redshift spaces:
\beq
\label{eq:mapping}
\bfs=\bfr+\frac{\bfv \cdot \hat{z}}{aH}\hat{z},
\eeq
where $\bfr$ and $\bfs$ denote position vectors in real and redshift spaces, respectively, and $\bfv$, $a$ and $H$ are the physical peculiar velocity, the scale factor and the Hubble parameter. Throughout the paper, we will work with the distant-observer limit, and choose $\hat{z}$ direction as the line-of-sight direction. Following the derivation of~\cite{Taruya10}, we begin by writing down a non-perturbative expression for redshift-space density power spectrum: 
\begin{equation}
P^{\rm(S)}(k,\mu)=\int d^3\bfx\,e^{i\,\bfk\cdot\bfx}
\bigl\langle e^{j_1A_1}A_2A_3\bigr\rangle\,, 
\label{eq:Pkred_exact}
\end{equation}
where we define
\begin{eqnarray}
&j_1= -i\,k\mu ,\nonumber\\
&A_1=u_z(\bfr)-u_z(\bfr'),\nonumber\\
&A_2=\delta(\bfr)+\,\nabla_zu_z(\bfr),\nonumber\\
&A_3=\delta(\bfr')+\,\nabla_zu_z(\bfr').\nonumber
\end{eqnarray}
Here, $\bfx=\bfr-\bfr'$, $\bfu\equiv-\bfv/(aH)$, and $u_z$ is the radial direction component of $\bfu$. The $\mu$ denotes the directional cosine of the angle between $\bfk$ and the line of sight. Eq.~(\ref{eq:Pkred_exact}) has been derived based on Eq.~(\ref{eq:mapping}). Thus, apart from the assumptions of distant-observer limit and single-stream flow in Eq.~(\ref{eq:mapping}), Eq.~(\ref{eq:Pkred_exact}) describes the non-perturbative nature of the mapping from real-space density clustering to redshift-space density clustering.

The pairwise velocity field, $A_1$, when expanded from the exponent, produces an indefinite series of higher-order polynomials, illustrating that nonlinear mapping induces non-perturbative non-Gaussian corrections in the two-point statistics. We rewrite the ensemble average $\langle e^{j_1A_1}A_2A_3\rangle$ in terms of the connected moments (cumulants) as~\cite{Taruya10}
\bea
\langle e^{j_1A_1}A_2A_3\rangle=
\exp \left\{\langle e^{j_1A_1}\rangle_c\right\}
\left[\langle e^{j_1A_1}A_2A_3 \rangle_c+ 
\langle e^{j_1A_1}A_2\rangle_c \langle e^{j_1A_1}A_3 \rangle_c \right]. \nonumber
\eea
Then Eq. (\ref{eq:Pkred_exact}) is recast as
\bea
&P^{\rm(S)}(k,\mu)=\int d^3\bfx \,\,e^{i\bfk\cdot\bfx}\,\,
\exp \left\{\langle e^{j_1A_1}\rangle_c\right\}
\left[\langle e^{j_1A_1}A_2A_3 \rangle_c+ 
\langle e^{j_1A_1}A_2\rangle_c \langle e^{j_1A_1}A_3 \rangle_c \right].
\label{eq:Pkred_exact2}
\eea
Note that the terms inside the brackets involve the squashing Kaiser effect, and when Taylor-expanded, the higher-order polynomials give either a mild enhancement or suppression of the power spectrum amplitude. On the other hand, in the presence of random velocity field, the prefactor, $\exp \left\{\langle e^{j_1A_1}\rangle_c\right\}$, always leads to a strong damping of the power spectrum, and is sensitively affected by small-scale virial motion. The so-called FoG effect arises from this prefactor.

In theoretically modeling redshift-space power spectrum, 
the spatial correlation in the exponential prefactor has been often ignored, and $\exp \left\{\langle e^{j_1A_1}\rangle_c\right\}$ is assumed to be independent of the separation vector $\bfx$
~\cite{Cole94,Vogeley94,Scoccimarro04,Taruya10}
\beq
P^{\rm (S)}(k,\mu)=D^{\rm FoG}(k\mu\sigma_z)P_{\rm perturbed}(k,\mu) \nonumber,
\label{eq:phenom}
\eeq
where $D^{\rm FoG}(k\mu\sigma_z)$ is the FoG term originated from the exponential prefactor, and we introduce the free parameter $\sigma_z$, which is related to the line-of-sight velocity dispersion through $\sigma_z^2\equiv\left\langle u_z^2\right\rangle_c$. The $P_{\rm perturbed}$ represents the Fourier transformation of the terms in the bracket in Eq.~(\ref{eq:Pkred_exact2}).

In this paper, on top of this modeling, we will examine an extension to include a part of spatial correlation in the exponential prefactor~\cite{Zhangrsd,Zheng13,Zheng16a}. This is done by formally writing Eq.~(\ref{eq:Pkred_exact2}) as
\bea
&P^{\rm(S)}(k,\mu)=D^{\rm FoG}_{\rm 1pt}(k\mu\sigma_z)\, \int d^3\bfx \,\,e^{i\bfk\cdot\bfx}D^{\rm FoG}_{\rm corr}(k\mu,\bfx)\left[\langle e^{j_1A_1}A_2A_3 \rangle_c+ 
\langle e^{j_1A_1}A_2\rangle_c \langle e^{j_1A_1}A_3 \rangle_c \right].
\label{eq:Pkred_exact3}
\eea
Here $D^{\rm FoG}_{\rm 1pt}(k\mu)$ is the same as defined above, and 
the factor $D^{\rm FoG}_{\rm corr}$ represents the exponential prefactor which includes the spatial correlation. In what follows, for a functional form of $D^{\rm FoG}_{\rm 1pt}(k\mu)$, we adopt the Gaussian form: 
\bea
D^{\rm FoG}_{\rm 1pt}(k\mu\sigma_z) = {\rm exp}\left[-\left(k\mu\sigma_z
\right)^2 \right]
\label{eq:RSD_model_extended}
\eea
with $\sigma_z$ being the free parameter describing the one--dimensional velocity dispersion.

In Eq.~(\ref{eq:Pkred_exact3}), while we need to keep the exponential prefactor $D^{\rm FoG}_{\rm 1pt}$ as non-perturbative damping term, the rest of the exponential factors, $e^{j_1A_1}$, may be expanded in powers of $j_1$, and this would be validated as long as we are interested in the small $j_1$. 
Collecting the terms at $\mathcal{O}(j_1^2)$ order, we have
\bea
&&D^{\rm FoG}_{\rm corr}(k\mu,\bfx)\left[\langle e^{j_1A_1}A_2A_3 \rangle_c+\langle e^{j_1A_1}A_2\rangle_c \langle e^{j_1A_1}A_3 \rangle_c \right] 
\simeq j_1^0\langle A_2A_3\rangle_c + j_1^1\langle A_1A_2A_3\rangle_c 
\nonumber\\
&&\qquad
+j_1^2\Bigl\{\langle A_1A_2\rangle_c\langle A_1A_3\rangle_c + \frac{1}{2}\,\langle A_1^2A_2A_3\rangle_c-\langle u_z u_z'\rangle_c\langle 
A_2A_3\rangle_c \Bigr\} +\mathcal{O}(j_1^3)\,.
\label{eq:expansion}
\eea
In the above, the zeroth-order term, $\langle A_2A_3\rangle_c$, corresponds to the squashing Kaiser term at linear order, and assuming the irrotational flow, this leads to the expression, $P_{\delta\delta}+2\mu^2P_{\delta\Theta}+\mu^4P_{\Theta\Theta}$, with $\Theta$ being the velocity-divergence field, $\Theta\equiv -\nabla\cdot \bfv/(aH)=\nabla \cdot \bfu$. The rest of the terms are regarded as higher-order corrections characterizing the nonlinear correlation between density and velocity fields, and substituted into Eq.~(\ref{eq:Pkred_exact3}), they produce the following corrections ~\cite{Taruya10,Taruya13}.
\begin{eqnarray}
  A(k,\mu)&=& j_1\,\int d^3\bfx \,\,e^{i\bfk\cdot\bfx}\,\,\langle A_1A_2A_3\rangle_c,\nonumber\\
  B(k,\mu)&=& j_1^2\,\int d^3\bfx \,\,e^{i\bfk\cdot\bfx}\,\,\langle A_1A_2\rangle_c\,\langle A_1A_3\rangle_c,\nonumber\\
  T(k,\mu)&=& \frac{1}{2} j_1^2\,\int d^3\bfx \,\,e^{i\bfk\cdot\bfx}\,\,\langle A_1^2A_2A_3\rangle_c,\nonumber \\
  F(k,\mu)&=& -j_1^2\,\int d^3\bfx \,\,e^{i\bfk\cdot\bfx}\,\,\langle u_z u_z'\rangle_c\langle A_2A_3\rangle_c, \nonumber
\end{eqnarray}
Thus, the model of redshift-space power spectrum examined in this paper is summarized as~\cite{Zheng16a} 
\bea
\label{eq:Pkred_final}
P^{\rm (S)}(k,\mu)&=&D^{\rm FoG}(k\mu\sigma_z)P_{\rm perturbed}(k,\mu) \\
&=&D^{\rm FoG}(k\mu\sigma_z)\,\Bigl[P_{\delta\delta}(k)+2\mu^2P_{\delta\Theta}(k)+\mu^4P_{\Theta\Theta}(k)  
\nn
\\
&&\qquad\qquad\qquad\qquad
+A(k,\mu)+B(k,\mu)+T(k,\mu)+F(k,\mu)\Bigr] \nonumber.
\eea
In Ref.~\cite{Zheng16a}, the validity of the truncation in Eq.~(\ref{eq:Pkred_final}) has been investigated, and based on the measurement of each correction term from the $N$-body simulations, the model is shown to give an accurate prediction for the 2D power spectrum at $k\lesssim0.2\hompc$. In what follows, varying cosmological models, we will further investigate the accuracy of the prediction based on Eq.~(\ref{eq:Pkred_final}), and study how well the model can be used for the theoretical template to accurately estimate the growth functions.

\begin{figure*}
\begin{center}
\resizebox{3.in}{!}{\includegraphics{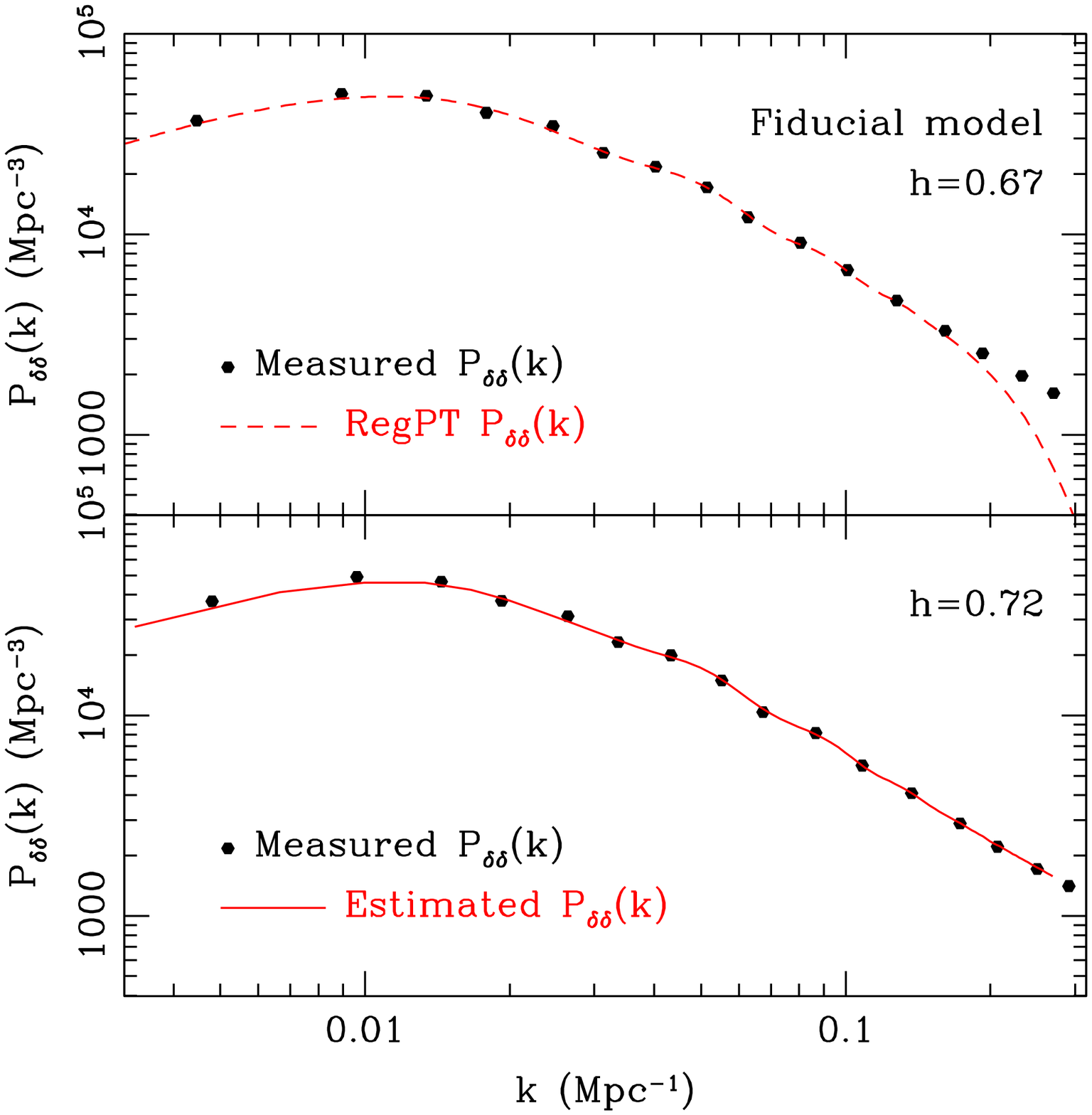}}\hfill
\resizebox{3.in}{!}{\includegraphics{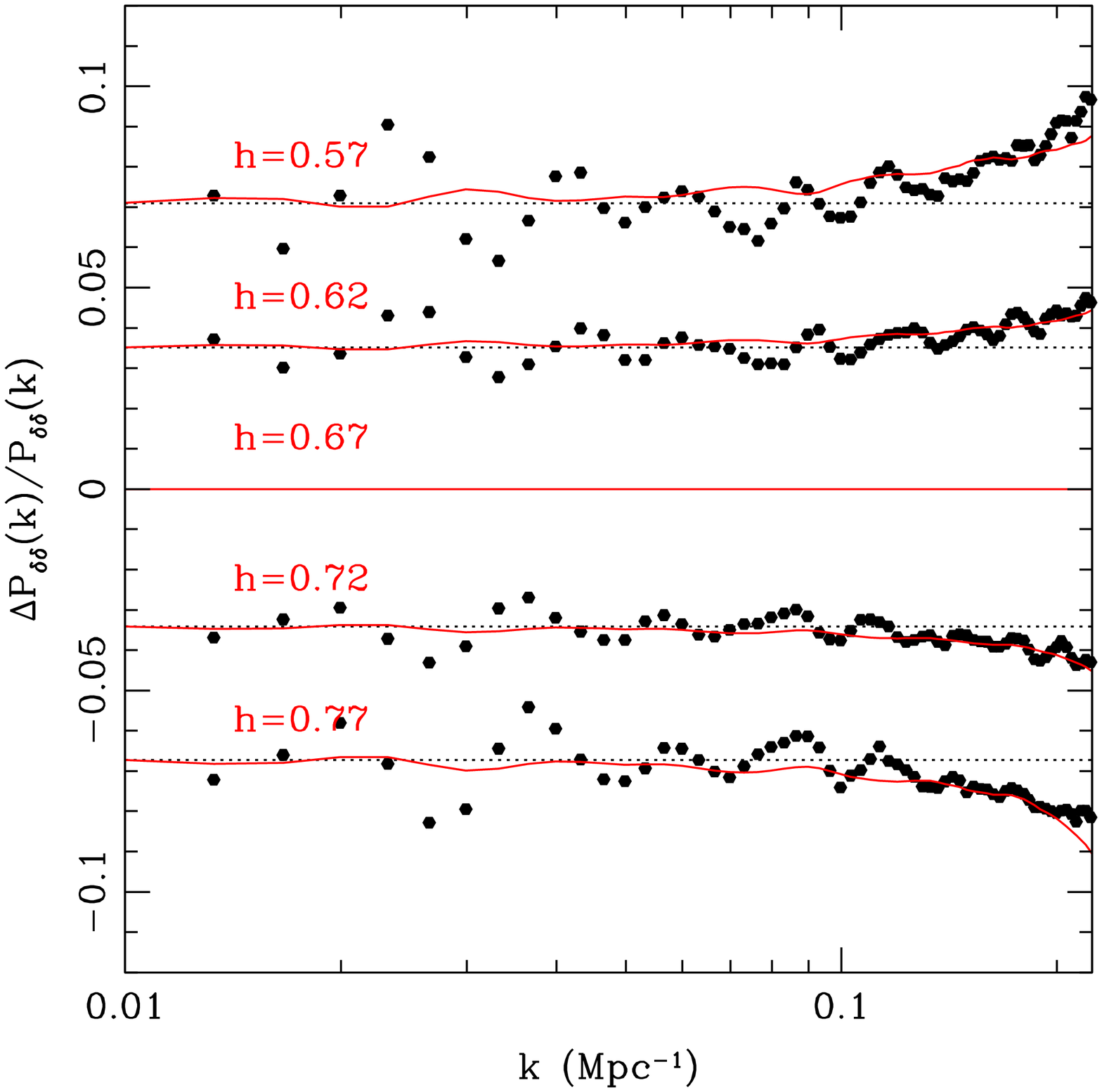}} 
\end{center}
\caption{{\it Left}: Broad-band shape of the power spectrum for density field at $z=0.5$. The results for the fiducial cosmological model with $h=0.67$ are plotted in top panel. The measured $\bar P_{\delta\delta}(k)$ and the theoretical $\bar P^{\rm th}_{\delta\delta}(k)$ are depicted as black dotted points and red dashed curve, respectively. The theoretical curve is computed with RegPT ~\cite{Taruya10,Taruya13}, assuming no systematics (i.e., $\bar{\cal O}^{(n)}\rightarrow 0$ limit). In bottom panel, we examine the case in another cosmology model with $h=0.72$. The measured $P_{\delta\delta}(k)$ for new cosmology model is depicted as black dotted points. The calibrated $P_{\delta\delta}(k)$ using Eq.~(\ref{eq:estimatedPk}) is plotted in red solid curve. {\it Right}: Fractional difference of the density power spectra at for cosmological models with different value of $h$. The results at $z=0.5$ are particularly shown. Black points represent the results in fiducial cosmology, while the results in other models with $(h=0.57,0.62,0.72,0.77)$ are shown from top to bottom. Red solid curves represent the results based on the calibrated power spectrum in Eq.~(\ref{eq:estimatedPk}) using the scaling relations (see text in Sec.~\ref{subsec:high-precison_PXY}).}
\label{fig:Pkdd}
\end{figure*}

\subsection{High-precision modeling of $P_{XY}(k)$}
\label{subsec:high-precison_PXY}

On large scales of our interest, where the gravitational clustering of matter/galaxy distribution is still in the weakly nonlinear regime, the perturbation theory (PT) treatment is supposed to work well, and one may use it as an accurate theoretical prediction. However, frequently used single-stream approximation when performing the PT calculations is shown to be very sensitive to the small-scale clustering to which the PT cannot be properly applied \cite{Blas_etal2013,Bernardeau_etal2012,Nishimichi:2014rra}. This is even true for the prediction at large-scales, and we must cure the UV sensitive behavior in the PT calculations (e.g., \cite{Baumann_etal2012,Carrasco_etal2012,Hertzberg2014,Baldauf:2015aha}). Further, goal of this paper is to exploit a method for a percent-level estimation of the growth functions. Even at large scales, higher-order corrections that are usually ignored in the prediction at few-percent level might play a role, leading to a non-negligible systematic error. Hence, in this paper, we consider a hybrid approach in which the uncertainty or systematics in the PT-based prediction is calibrated and corrected with $N$-body simulations. For this purpose, we measured accurately the power spectrum using $100$ simulations with the box size of $L=1.89\,h^{-1}$Gpc and with the number of particles $N=1024^3$ \cite{Zheng16c} in the fiducial cosmology with $h=0.67$, but we are much interested in extending an accurate template in a given fiducial cosmology to those in other cosmology models. We additionally generate $4$ simulations, each of which has a unique $h=(0.57,0.62,0.72,0.77)$, in order to test the accuracy of our estimated power spectrum for different cosmologies.

First consider the power spectrum in a fiducial cosmology. We denote it by $\bar P_{XY}(k,z)$ $(X,Y=\delta\,\,\mbox{or}\,\,\Theta)$. To compute the power spectrum by PT, we adopt the RegPT treatment proposed by
~\cite{Taruya10,Taruya13}. This treatment is based on the resummed PT expansion referred to as the multi-point propagator expansion \cite{Bernardeau:2008fa}, in which all the statistical quantities including power spectrum are expanded in terms of the multi-point propagators. The expression of the power spectrum, valid at two-loop order (i.e., next--to--next--to--leading order) is given by 
\ba
&&\bar P_{XY}(k,z) = \bar \Gamma_X^{(1)}(k,z)\bar \Gamma_Y^{(1)}(k,z)\bar P^{i}(k)
\nn
\\
&&\quad
+2\int\frac{d^3\vec q}{(2\pi)^3}\bar \Gamma_X^{(2)}(\vec q, \vec k-\vec q,z) \bar \Gamma_Y^{(2)}(\vec q, \vec k-\vec q,z) 
\bar P^i(q) \bar P^{i}(|\vec k-\vec q|) 
\nn
\\
&&\quad
+6\int\frac{d^3\vec pd^3\vec q}{(2\pi)^6} \bar \Gamma_X^{(3)}(\vec p,\vec q, \vec k-\vec p-\vec q,z)\bar \Gamma_Y^{(3)}(\vec p,\vec q, \vec k-\vec p-\vec q,z) \bar P^i(p) \bar P^i(q) \bar P^i(|\vec k-\vec p-\vec q|), 
\nn
\\
&&\qquad\qquad (X,Y=\delta\,\,\mbox{or}\,\,\Theta).
\label{eq:pk_RegPT}
\ea
Here, barred quantities are those computed in the fiducial cosmological model, and $\bar P^i$ is the initial power spectrum. The function $\Gamma_X^{(n)}$ is the $(n+1)$-point propagator. In RegPT treatment, the propagators are constructed with standard PT calculations. While the standard PT is usually applied to a limited range of wavenumber, incorporating the result of a partial resummation in the high-$k$ limit, a regularized prediction of the propagators is applicable to a larger $k$ \cite{Bernardeau:2011dp}. 

Let us see the two-point propagator $\bar\Gamma_X^{(1)}$. The expression relevant at two-loop order is summarized as
\ba\label{eq:gammax1}
\bar \Gamma_X^{(1)}(k,z) = {\rm exp}\left(-\bar G_\delta^2\,\bar\gamma \right)\sum_n \bar G_X \bar G_\delta^{n-1} \bar{\cal C}^{(1)}_n(\bar \gamma),\qquad\quad(X=\delta\,\,\mbox{or}\,\,\Theta).
\ea
Here,  $\bar\gamma$ is defined by $\bar\gamma=k^2\bar \sigma^2_{\rm d}/2$ with $\bar \sigma_{\rm d}^2$ being the dispersion of displacement field. The $\sigma_{\rm d}$ is computed with the initial power spectrum through\footnote{Note that choice of the upper bound of the integral has been specified in
\cite{RegPT} in somewhat phenomenological way,  and there might be a possible uncertainty. However, it can be absorbed into $\bar{\cal O}$ in our prescription at least perturbatively.} $\bar \sigma_{\rm d}^2=\int_0^{k/2} (dq/6\pi^2)\bar P^i(q)$. The coefficients $\bar{\cal C}^{(n)}$ in Eq.~(\ref{eq:gammax1}) are expressed in terms of the standard PT results, and including the theoretical uncertainties, they are given by
\ba
\bar{\cal C}^{(1)}_1(\bar \gamma) &=&1, \\
\bar{\cal C}^{(1)}_3(\bar \gamma) &=& \bar \gamma + \bar \Gamma_{X,{\rm 1-loop}}^{(1)}(k),\\
\bar{\cal C}^{(1)}_5(\bar \gamma) &=& \bar \gamma^2/2 + \bar \gamma\bar \Gamma_{X,{\rm 1-loop}}^{(1)}(k) + \bar \Gamma_{X,{\rm 2-loop}}^{(1)}(k)+\bar{\cal O}^{(1)}_{X,5},\\
\bar{\cal C}^{(1)}_n(\bar \gamma) &=& \bar{\cal O}^{(1)}_{X,n},
\ea
and $\bar {\cal C}^{(1)}_n=0$ for even number of $n$. The $\bar G_X$ denotes the density ($X=\delta$) and velocity ($X=\Theta$) growth functions for the fiducial cosmology at the redshift $z$. These growth functions are the key quantities to be estimated from observations accurately and precisely, and are related to the linear growth factor $D_+$ and linear growth rate $f$ defined by $f\equiv d\ln D_+/d\ln a$ through $G_\delta=D_+$ and $G_\theta=f\,D_+$. The function $\Gamma_{X,{\rm n-loop}}^{(p)}$ represents the standard PT $(p+1)$-point propagator at $n$-loop order, whose explicit expression is given in
~\cite{RegPT,Bernardeau14}. The quantity $\bar{\cal O}^{(1)}_{X,n}$ characterizes the uncertainties or systematics in PT, which will be later calibrated with $N$-body simulations. We assume that the uncertainties arise not only from higher-order (three-loop) but also from two-loop order, partly due to the UV sensitive behavior of the single-stream PT calculation.

Similarly, the expression of the three-point propagator, $\bar \Gamma_X^{(2)}(k,z)$ is given by
\ba
\bar \Gamma_X^{(2)}(k,z) &=& {\rm exp}\left(-\bar G_\delta^2\,\bar\gamma \right)\sum_n \bar G_X\,\bar G_\delta^{n-1} \bar{\cal C}^{(2)}_n\,\,; \\
\bar{\cal C}^{(2)}_2(\bar \gamma) &=& \bar F_X^{(2)}(\vec q, \vec k-\vec q),\\
\bar{\cal C}^{(2)}_4(\bar \gamma) &=& \frac{\bar\gamma}{2} \bar F_X^{(2)}(\vec q, \vec k-\vec q)+\bar\Gamma_{X,{\rm 1-loop}}^{(2)}(\vec q, \vec k-\vec q)+\bar{\cal O}^{(2)}_{X,4}, \\ 
\bar{\cal C}^{(2)}_n(\bar \gamma) &=& \bar{\cal O}^{(2)}_{X,n}
\ea
and $\bar{\cal C}^{(2)}_n=0$ for odd number of $n$. Also, the expression of the four-point propagator, $\bar \Gamma_X^{(3)}(k,z)$, relevant at two-loop order, is
\ba
\bar \Gamma_X^{(3)}(k,z) &=& {\rm exp}\left(-\bar G_\delta^2\,\gamma \right)\sum_n \bar G_X\,\bar G_\delta^{n-1} \bar{\cal C}^{(3)}_n\,\,; \\
\bar{\cal C}^{(3)}_3(\bar \gamma)&=&\bar F_X^{(3)}(\vec p,\vec q, \vec k-\vec p-\vec q) + \bar{\cal O}^{(3)}_{X,3}, \\
\bar{\cal C}^{(3)}_n(\bar \gamma) &=& \bar{\cal O}^{(3)}_{X,n}.
\ea
Note that $\bar{\cal O}^{(2)}$ and $\bar{\cal O}^{(3)}$ represent the possible uncertainties.

\begin{figure}
\begin{center}
\resizebox{3.2in}{!}{\includegraphics{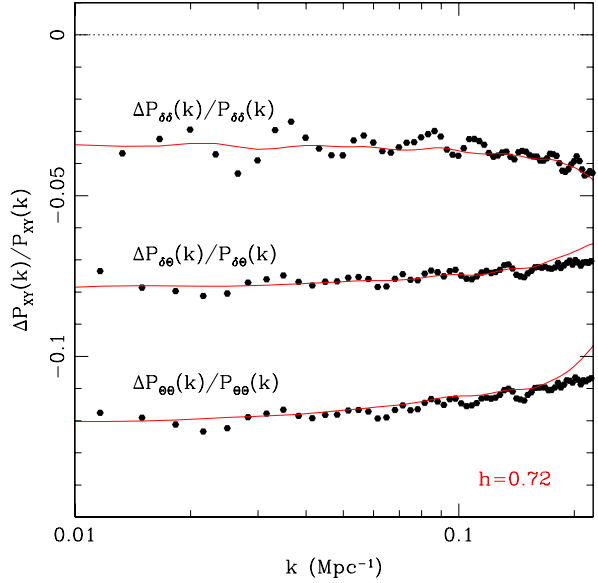}}
\end{center}
\vspace*{1.5cm}
\caption{Fractional difference between the measured $\bar P_{XY}(k)$ and the predictions based on the calibrated $\bar P^{\rm th}_{XY}(k)$ using scaling relations. The results are shown at $z=0.5$ for the cosmological model with $h=0.72$, which differs from $h=0.67$ in fiducial cosmological model. From top to bottom, we plot $\Delta P_{\delta\delta}$, $\Delta P_{\delta\Theta}$ and $\Delta P_{\Theta\Theta}$. }
\label{fig:dPkXY}
\end{figure}


Fig.~\ref{fig:Pkdd} compares the predicted density power spectrum with measured results from simulations at $z=0.5$. The dashed curve in the top--left panel represents the prediction based on Eq.~(\ref{eq:pk_RegPT}), setting $\bar{\cal O}^{(m)}_{X,n}\rightarrow 0$, which agrees very well with simulation at $k\lesssim 0.1\ompc$. At $k\gtrsim0.1\ompc$, discrepancies are manifest, and the predicted amplitude by RegPT rapidly falls off. While this is partly due to the exponential factor, $\exp(-\gamma)$, in the multi-point propagator, a lack of higher-order terms as well as a small systematics in the PT calculations can sensitively affect the high-$k$ prediction. Here, we characterize the difference between the measured and predicted power spectra by $\bar P_{XY}^{\rm res}$. We then divide the power spectrum into two pieces:
\ba
\bar P_{XY}(k,z)=\bar P^{\rm th}_{XY}(k,z)+\bar P^{\rm res}_{XY}(k,z),
\label{eq:pk_hybrid_fiducial}
\ea
where $\bar P^{\rm th}_{XY}$ represents the PT prediction with $\bar{\cal O}^{(m)}_{X,n}\rightarrow 0$. Collecting all the uncertainties introduced in the multi-point propagators, the residual power spectrum $\bar P_{XY}^{\rm res}$ is schematically expressed as
\ba
&&\bar P^{\rm res}_{XY}=\bar G_X \bar G_Y \bar G_\delta^4\,\Biggl\{
\left[{\cal O}^{(1)}_{Y,5} + {\rm higher} \right]\bar P^{i}
+\left[\bar{\cal O}^{(1)}_{X,5} + {\rm higher} \right]\bar P^{i},\nn\\
&&+\int \left[\bar{\cal O}^{(2)}_{Y,4}\bar F_Y^{(2)}+ {\rm higher}\right]\bar P^{i}\bar P^{i}
+\int \left[\bar{\cal O}^{(2)}_{X,4}\bar F_X^{(2)}+ {\rm higher}\right]\bar P^{i}\bar P^{i}, \nn\\
&&+\int\int \left[ \bar{\cal O}^{(3)}_{Y,3}\bar F_Y^{(3)}+ {\rm higher}\right]\bar P^{i}\bar P^{i}\bar P^{i} 
+\bar \int\int \left[\bar{\cal O}^{(3)}_{X,3}\bar F_X^{(3)}+ {\rm higher}\right]\bar P^{i}\bar P^{i}\bar P^{i} \Biggr\}.
\label{eq:P_res_fid}
\ea
Here, the uncertainty $\bar{\cal O}^{(m)}_{X,n}$ is assumed to be small, and to be perturbatively treated. The expression implies that apart from a detailed scale-dependent behavior, time dependence is characterized by 
$G_XG_YG_\delta^4$. Thus, once we calibrate the $\bar P_{XY}^{\rm res}$ at a given redshift, we may use it for the prediction at another redshift by simply rescaling the calibrated residuals. Further, for cosmological models close to the fiducial model, the scale dependence of the higher-order PT corrections is generally insensitive to the cosmology, and we may also apply the calibrated $\bar P_{XY}^{\rm res}$ to other cosmological models.

To check if the scaling ansatz works well, we write the power spectrum in the cosmological model different from fiducial one as
\ba\label{eq:estimatedPk}
P_{XY}(k) = P^{\rm th}_{XY}(k,z)+ P^{\rm res}_{XY}(k,z),
\ea
where the power spectrum $P^{\rm th}_{XY}(k,z)$ is evaluated based on the PT predictions in the fiducial model, given at Eq.~(\ref{eq:pk_RegPT}), but the multi-point propagators are replaced with the rescaled one:
\ba
\Gamma_X^{(m)} &=& {\rm exp}\left(-G_X^2\bar \gamma \right)\sum_n G_X\,G_\delta^{n-1} \bar{\cal C}^{(m)}_n(\gamma),\qquad\quad (m=1,2,3)
\nn
\ea
with $\bar{\cal O}^{(m)}_{X,n}\rightarrow 0$. The $\gamma_X$ is the rescaled version of $\bar\gamma$, defined by,
\ba
\gamma_X\equiv \left(\frac{G_\delta}{\bar G_\delta G_X}\right)^2\bar\gamma.
\ea
On the other hand, the residual power spectrum, $P^{\rm res}_{XY}$, is simply evaluated with the calibrated result in the fiducial model, $\bar P^{\rm res}_{XY} $ through
\ba
P^{\rm res}_{XY}=\left(\frac{G_X\,G_Y\,G_\delta^4}{\bar G_X \bar G_Y\,\bar G_\delta^4}\right) \bar P^{\rm res}_{XY},\qquad 
(XY=\delta\delta,\,\,\delta\Theta,\,\mbox{and}\,\,\Theta\Theta). 
\ea


Bottom left panel of Fig.~\ref{fig:Pkdd} shows the calibrated PT power spectrum using Eq.~(\ref{eq:estimatedPk}) in the cosmological model with $h=0.72$, depicted as red solid curve. The calibrated power spectrum matches well with the measured result in $N$-body simulations at $k\lesssim0.3\ompc$. We then use Eq.~(\ref{eq:estimatedPk}) to predict the density power spectrum in cosmological models which have different Hubble parameters. The results in the models with $h=(0.57,0.62,0.72,0.77)$ are plotted in right panels of Fig.~\ref{fig:Pkdd} (from top to bottom), shown as the fractional difference, $[P_{\delta\delta}-\bar P_{\delta\delta}]/\bar P_{\delta\delta}$.  For reference, we also plot in dotted horizontal lines the simple scaling factor predicted by linear theory, i.e., $(G_\delta/\bar G_\delta)^2-1$. Up to $k~\sim 0.1\ompc$, the fractional difference obtained from measured power spectra is consistent with linear theory prediction, but it exhibits a non--trivial scale-dependence at $k\ga 0.1\ompc$. The prediction based on the calibrated PT power spectrum in the fiducial model fairly traces the measured results at $k\lesssim 0.2\ompc$, indicating that our scaling treatment is successful. In Fig.~\ref{fig:dPkXY}, the scaling treatment is further tested for $P_{\delta\Theta}$ and $P_{\Theta\Theta}$ in the fiducial model. The predicted power spectra are pretty much consistent with measured results of power spectrum difference at $k\lesssim 0.2\ompc$.

\begin{figure*}
\begin{center}
\resizebox{2.0in}{!}{\includegraphics{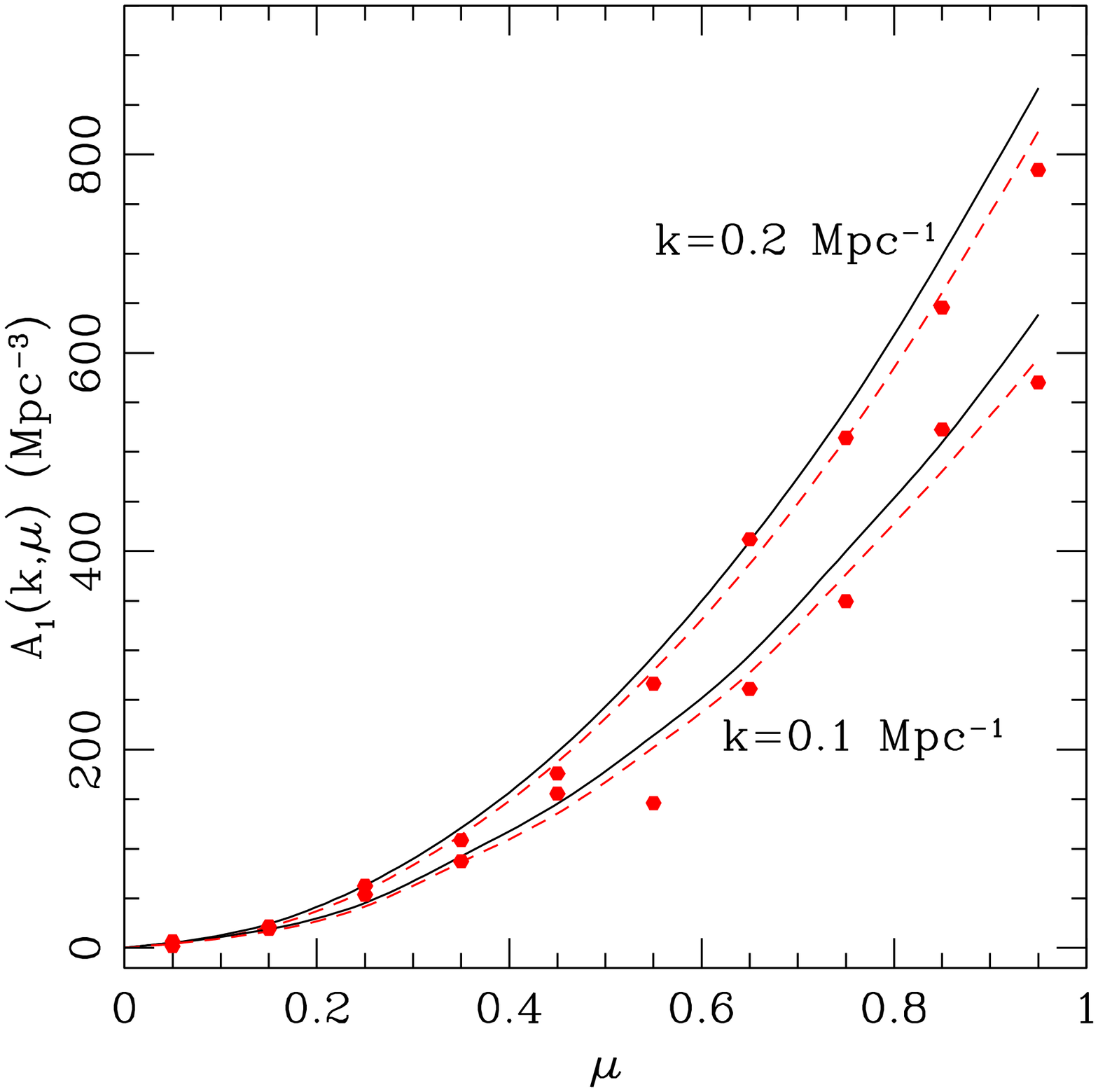}}\hfill
\resizebox{2.0in}{!}{\includegraphics{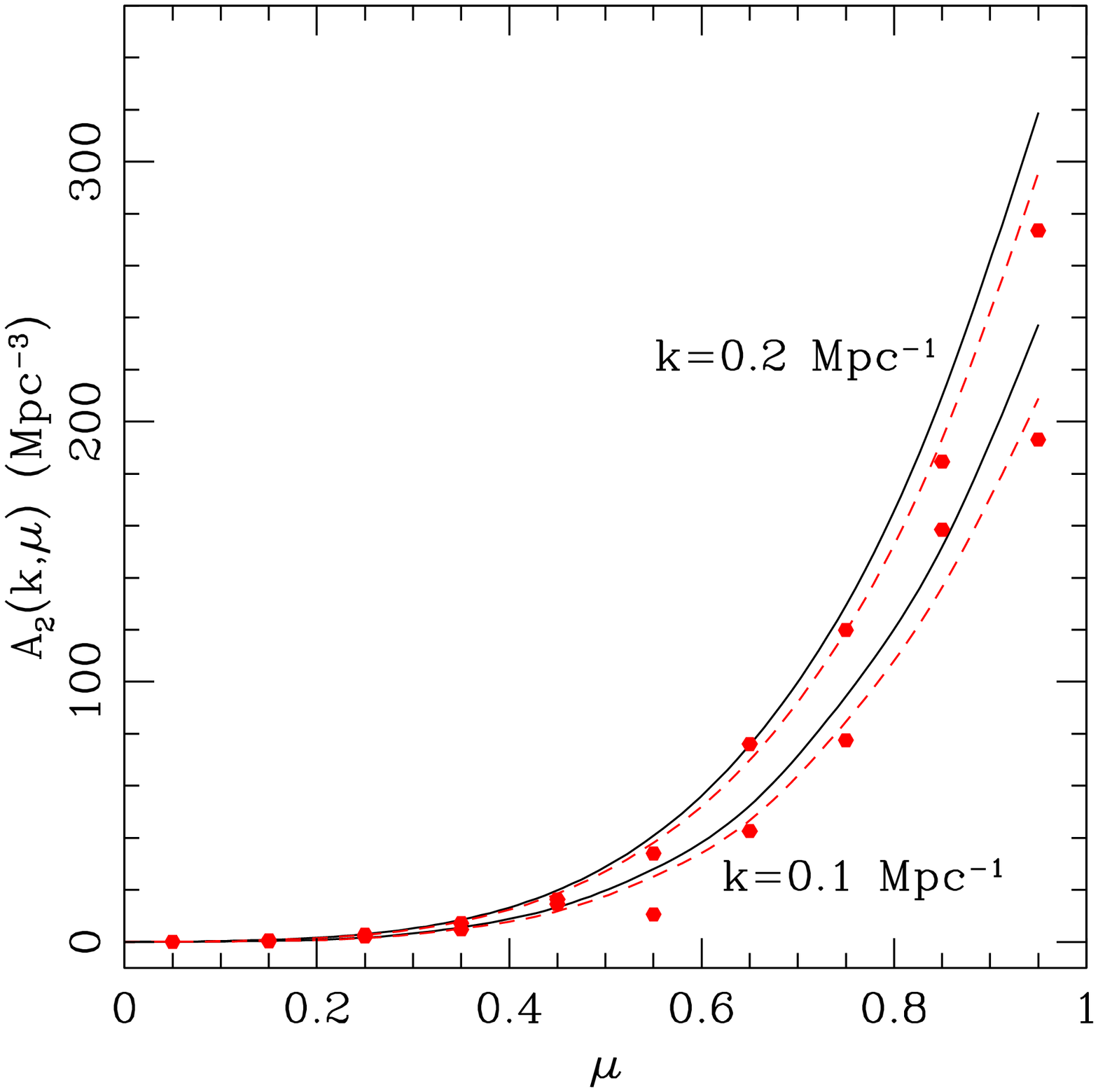}} \hfill
\resizebox{2.0in}{!}{\includegraphics{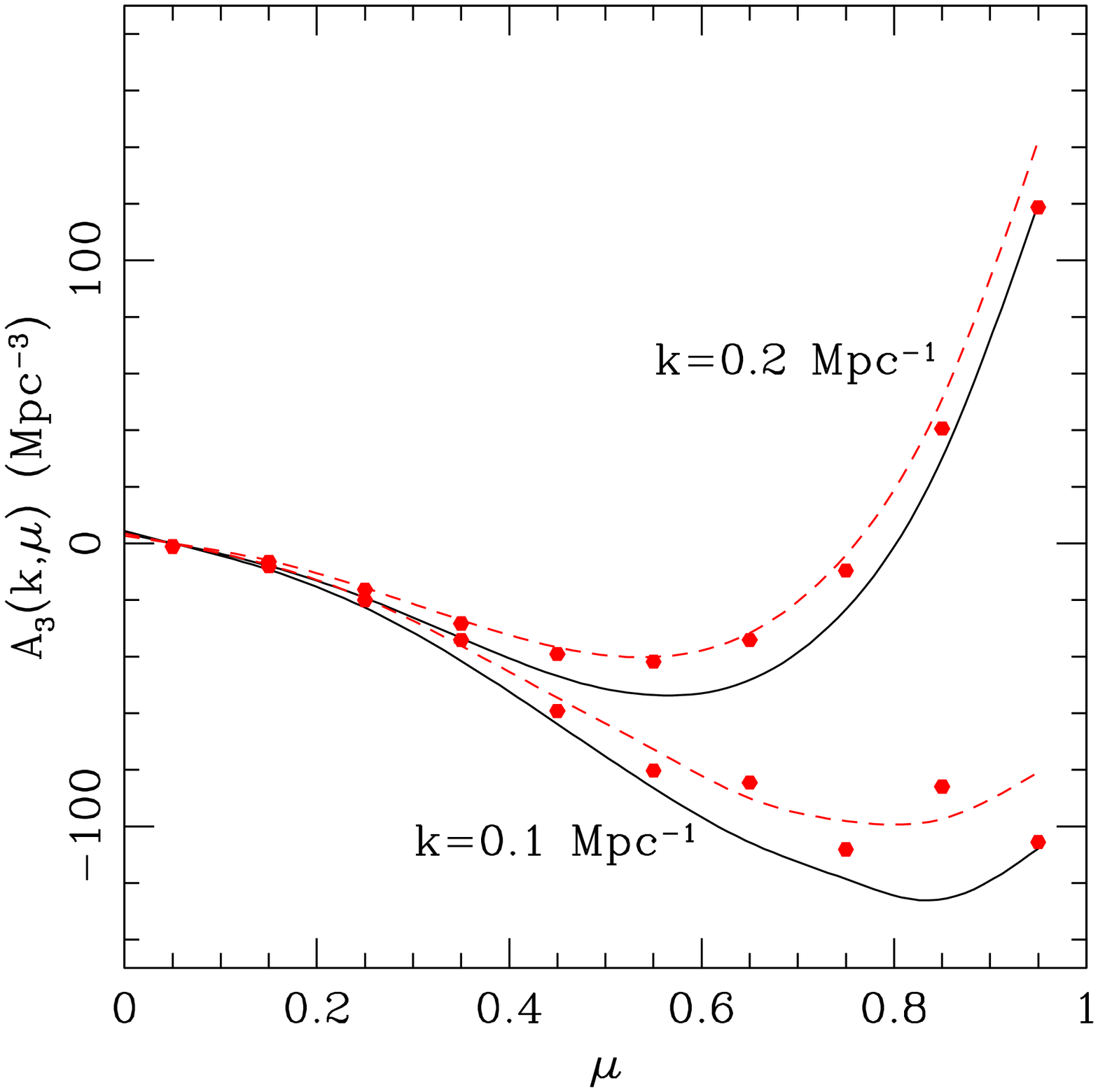}} \\
\resizebox{2.0in}{!}{\includegraphics{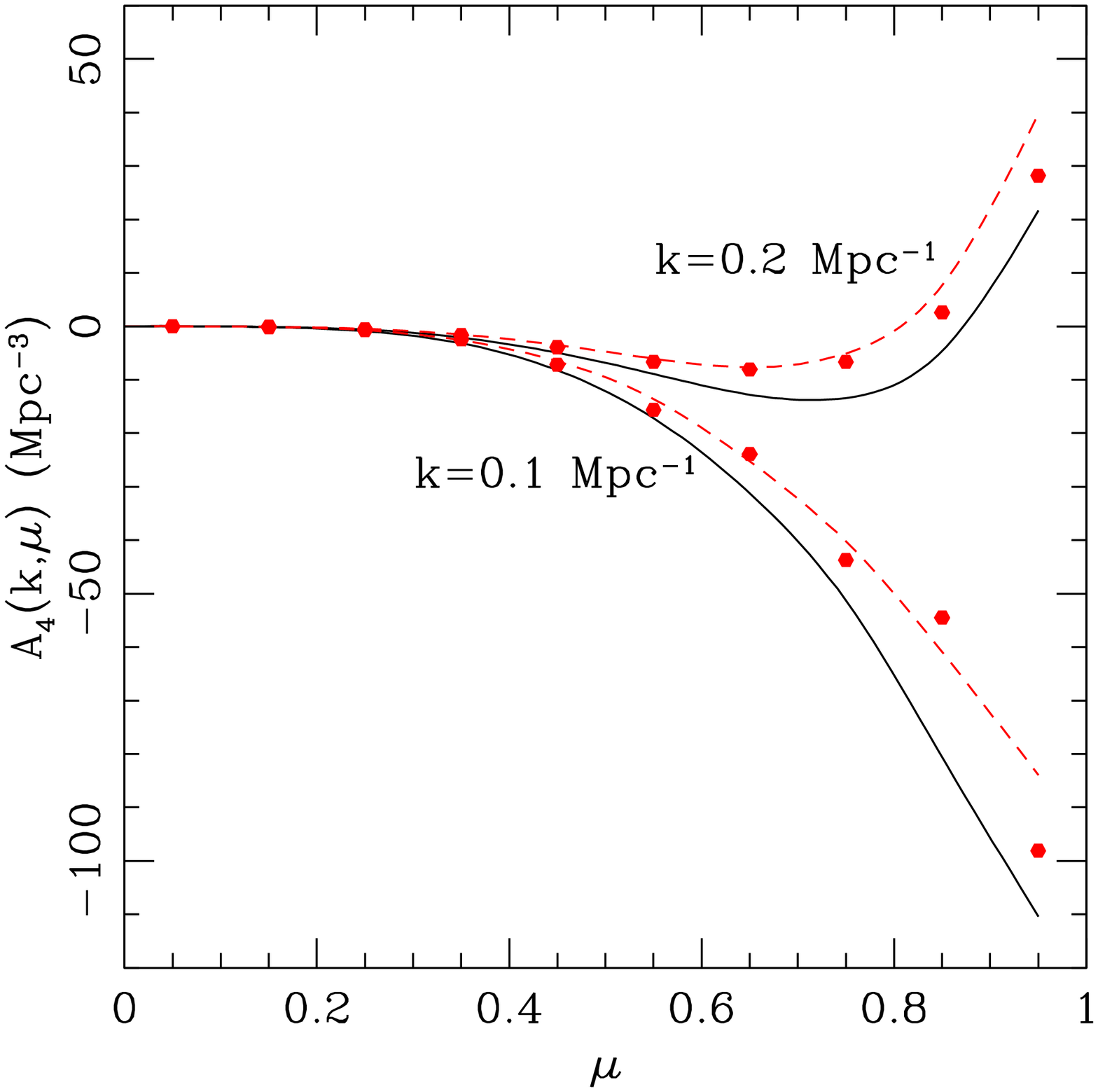}}\hfill
\resizebox{2.0in}{!}{\includegraphics{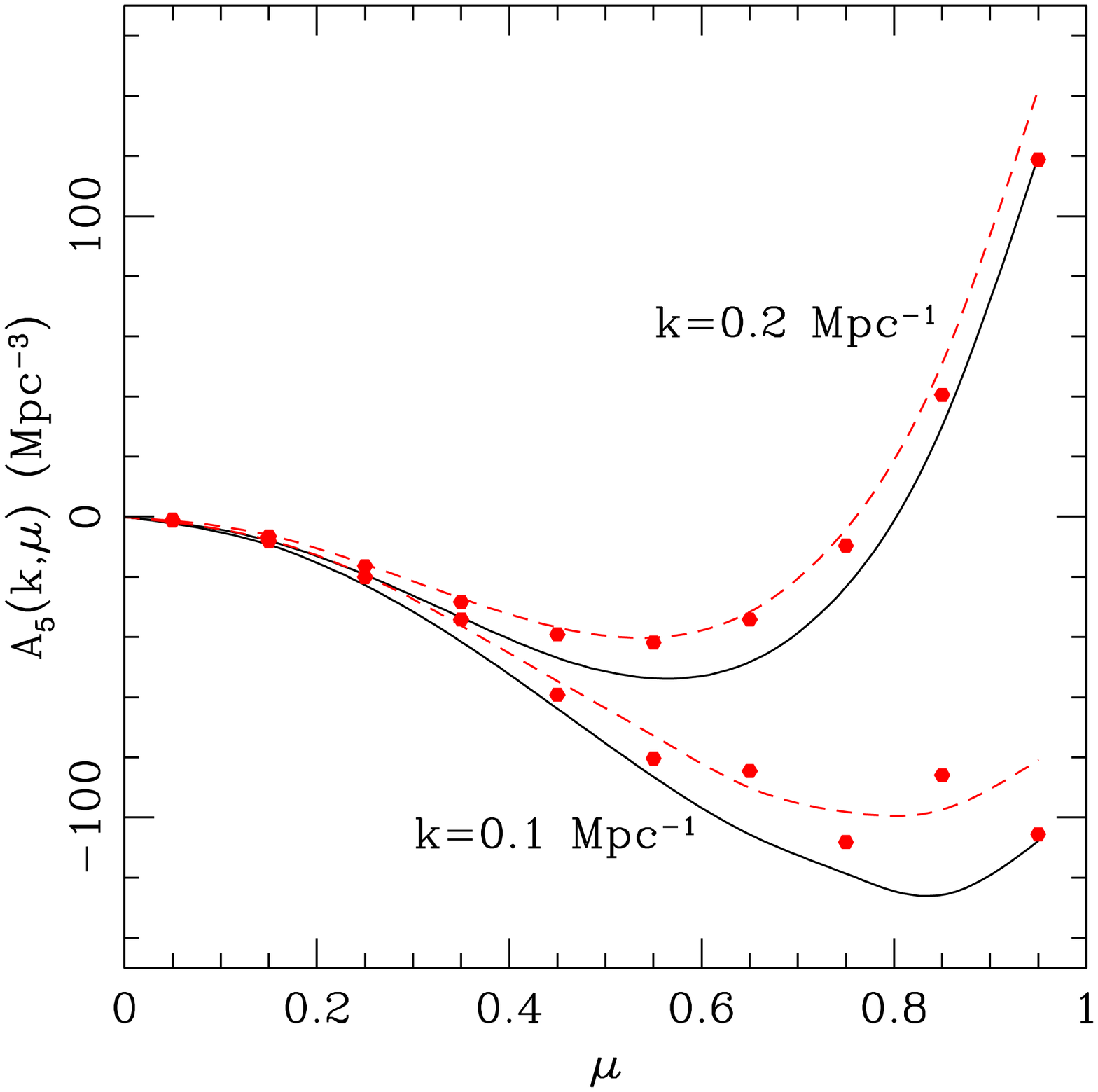}} \hfill
\resizebox{2.0in}{!}{\includegraphics{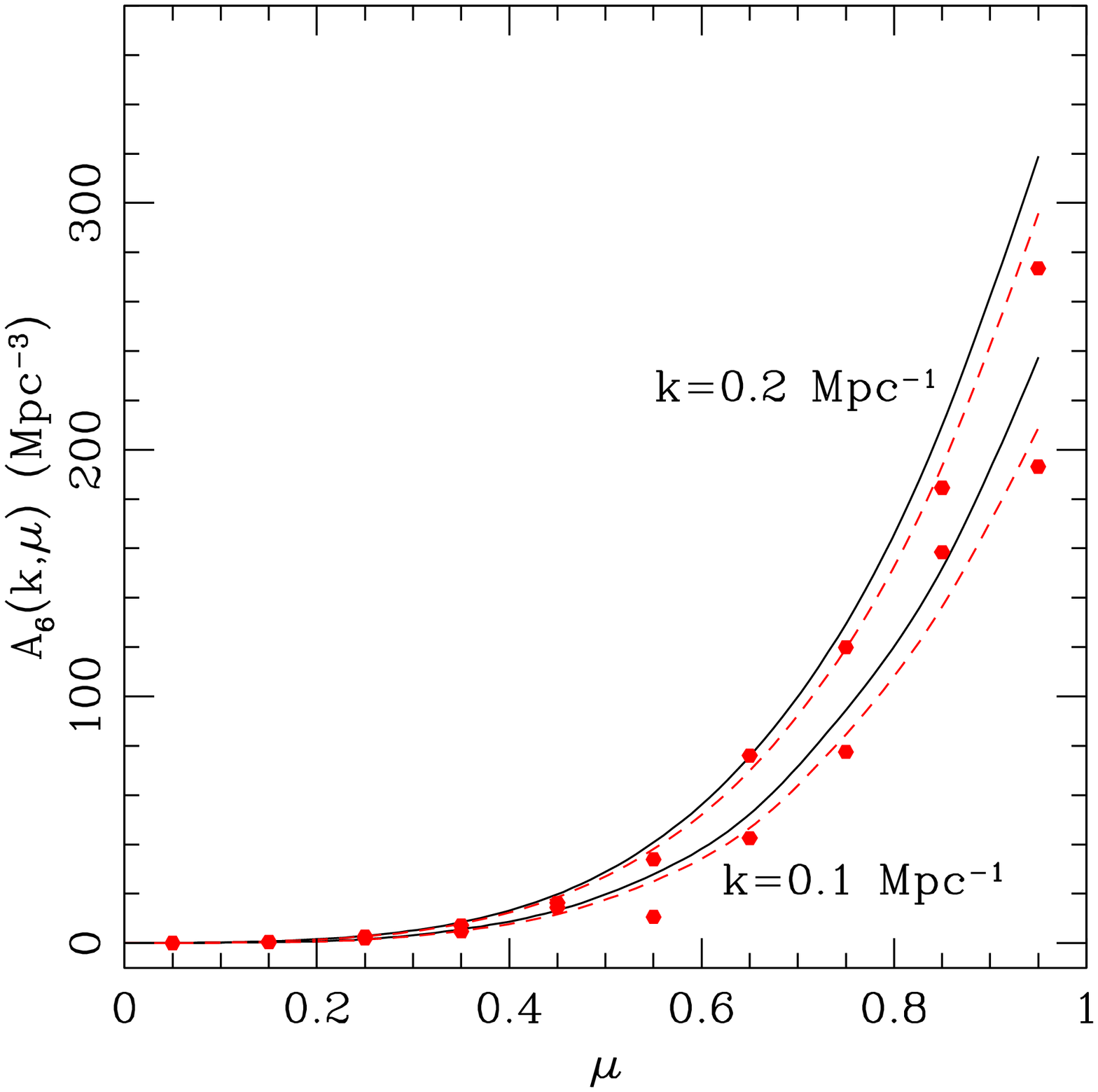}} 
\end{center}
\vspace*{1.0cm}
\caption{Higher-order correction in the redshift-space power spectrum, $A(k,\mu)$, at $z=0.5$. Dividing the function $A$ into six pieces [see Eqs.~(\ref{eq:A_term}) and (\ref{eq:A_1})-(\ref{eq:A_6})], the measured $\bar A_n$ are plotted in black solid curves for the fiducial cosmology. The measured $A_n$ is also shown in red dotted points for another cosmological model with $h=0.72$  at the specific scales of $k=0.1$ and $0.2\ompc$. The red dashed curves are the predictions based on the measured $\bar A_n$ in fiducial model using the scaling relation at Eq.~(\ref{eq:estimatedAn}). }
\label{fig:Akmu}
\end{figure*}

\subsection{High-precision modeling of higher-order corrections}

\begin{figure*}
\begin{center}
\resizebox{2.0in}{!}{\includegraphics{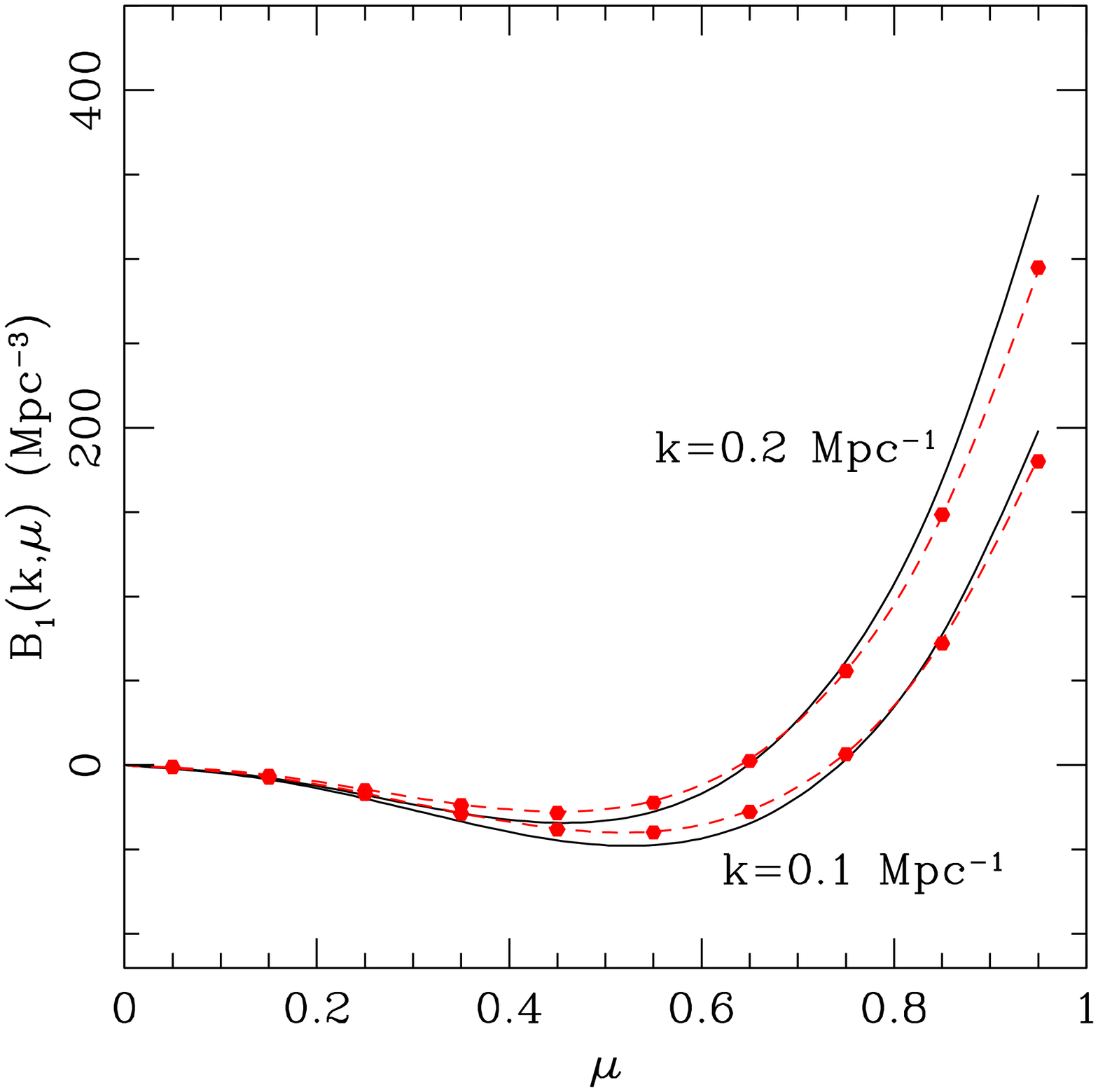}}
\resizebox{2.0in}{!}{\includegraphics{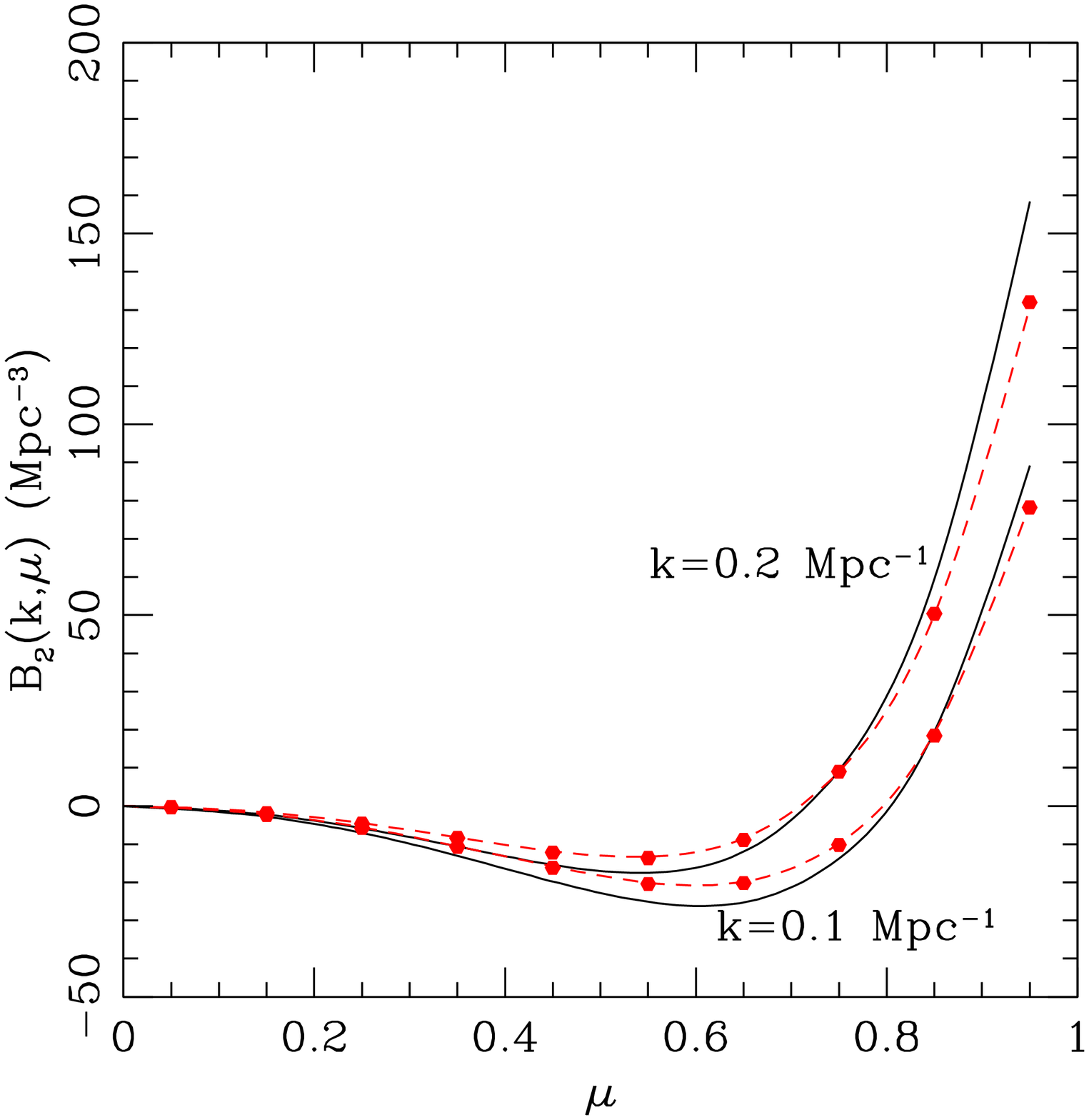}} \\
\resizebox{2.0in}{!}{\includegraphics{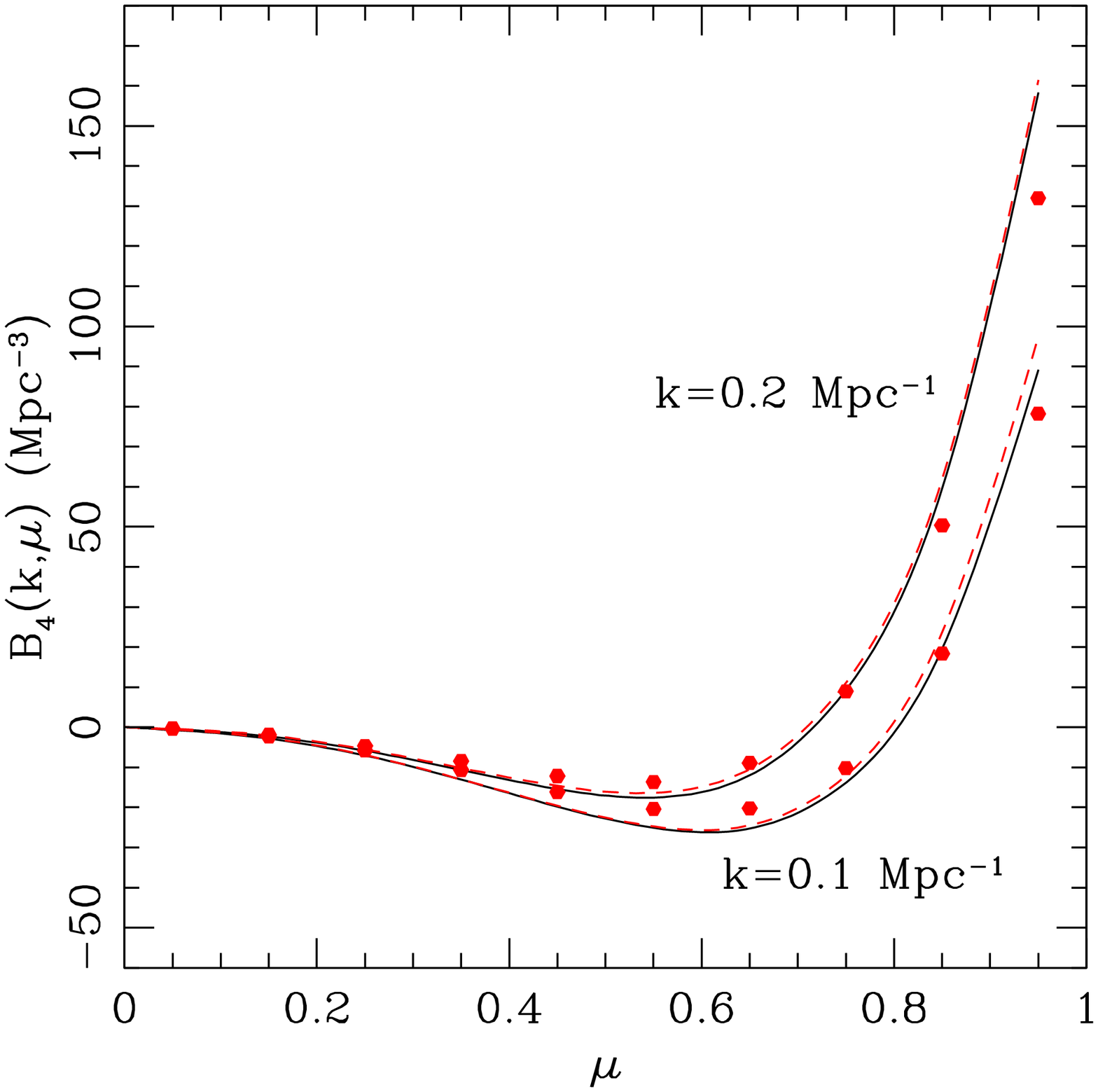}}
\resizebox{2.0in}{!}{\includegraphics{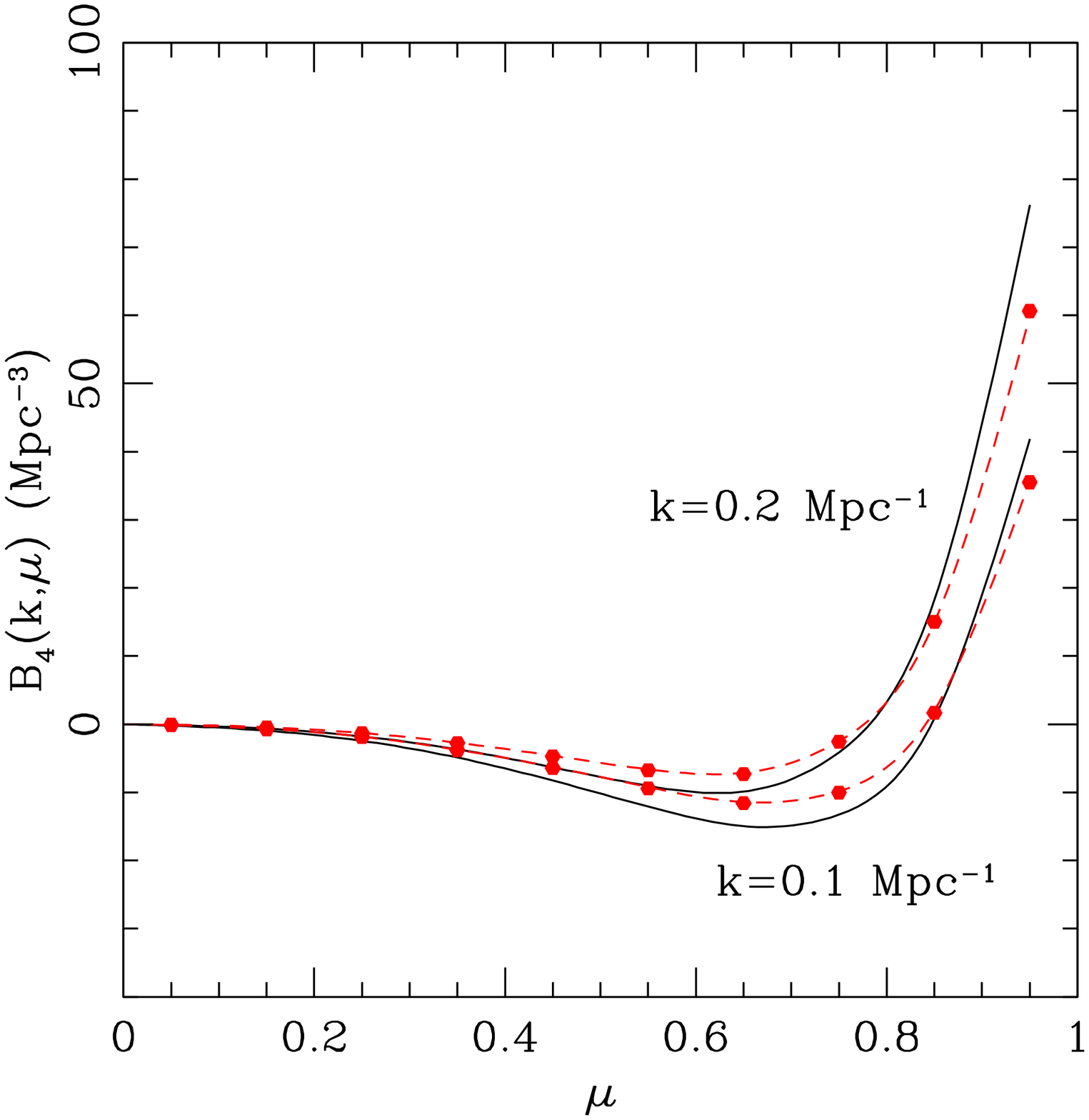}}
\end{center}
\vspace*{1.0cm}
\caption{Higher-order correction in the redshift-space power spectrum, $B(k,\mu)$, at $z=0.5$. Dividing the function $B$ into four pieces [see Eq.~(\ref{eq:estimatedBn}) and Appendix \ref{app:explicit_form_B_F_T}], the measured $\bar B_n$ are plotted in black solid curves for the fiducial cosmology. The measured $B_n$ is also shown in red dotted points for another cosmological model with $h=0.72$  at the specific scales of $k=0.1$ and $0.2\ompc$. The red dashed curves are the predictions based on the measured $\bar B_n$ in fiducial model using the scaling relation at Eq.~(\ref{eq:estimatedBn}). }
\label{fig:Bkmu}
\end{figure*}
\begin{figure*}
\begin{center}
\resizebox{2.0in}{!}{\includegraphics{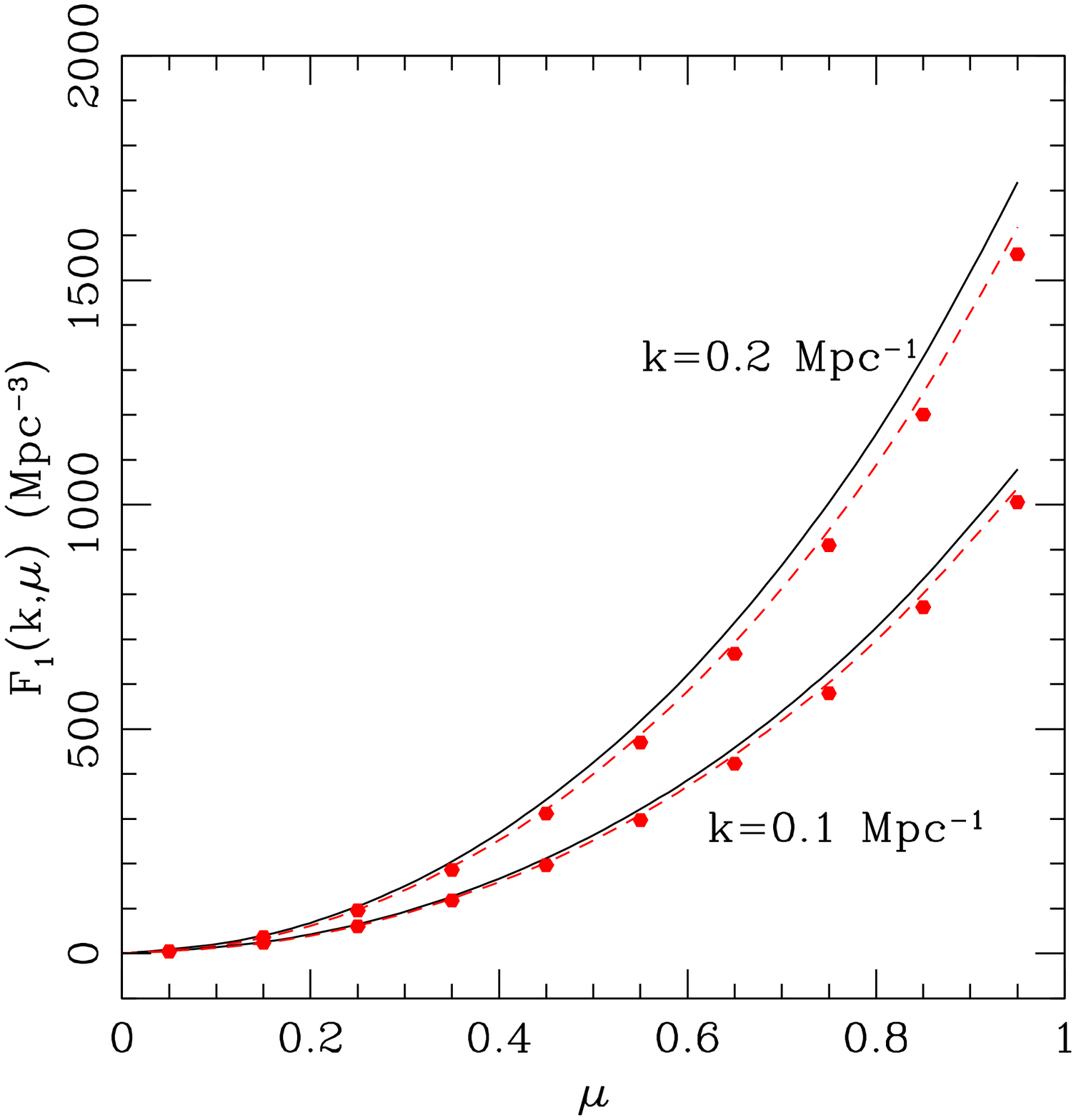}}\hfill
\resizebox{2.0in}{!}{\includegraphics{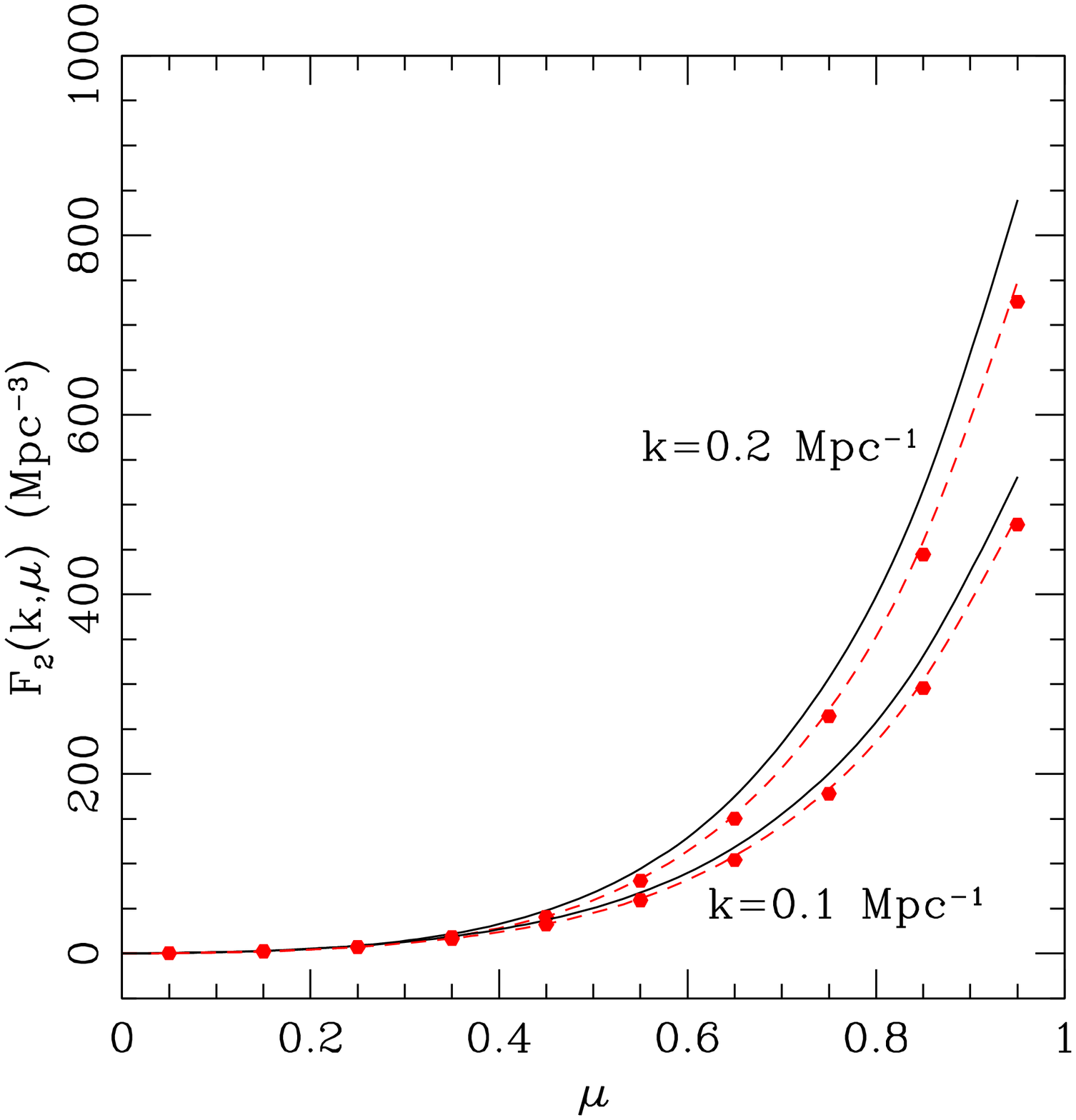}}\hfill
\resizebox{2.0in}{!}{\includegraphics{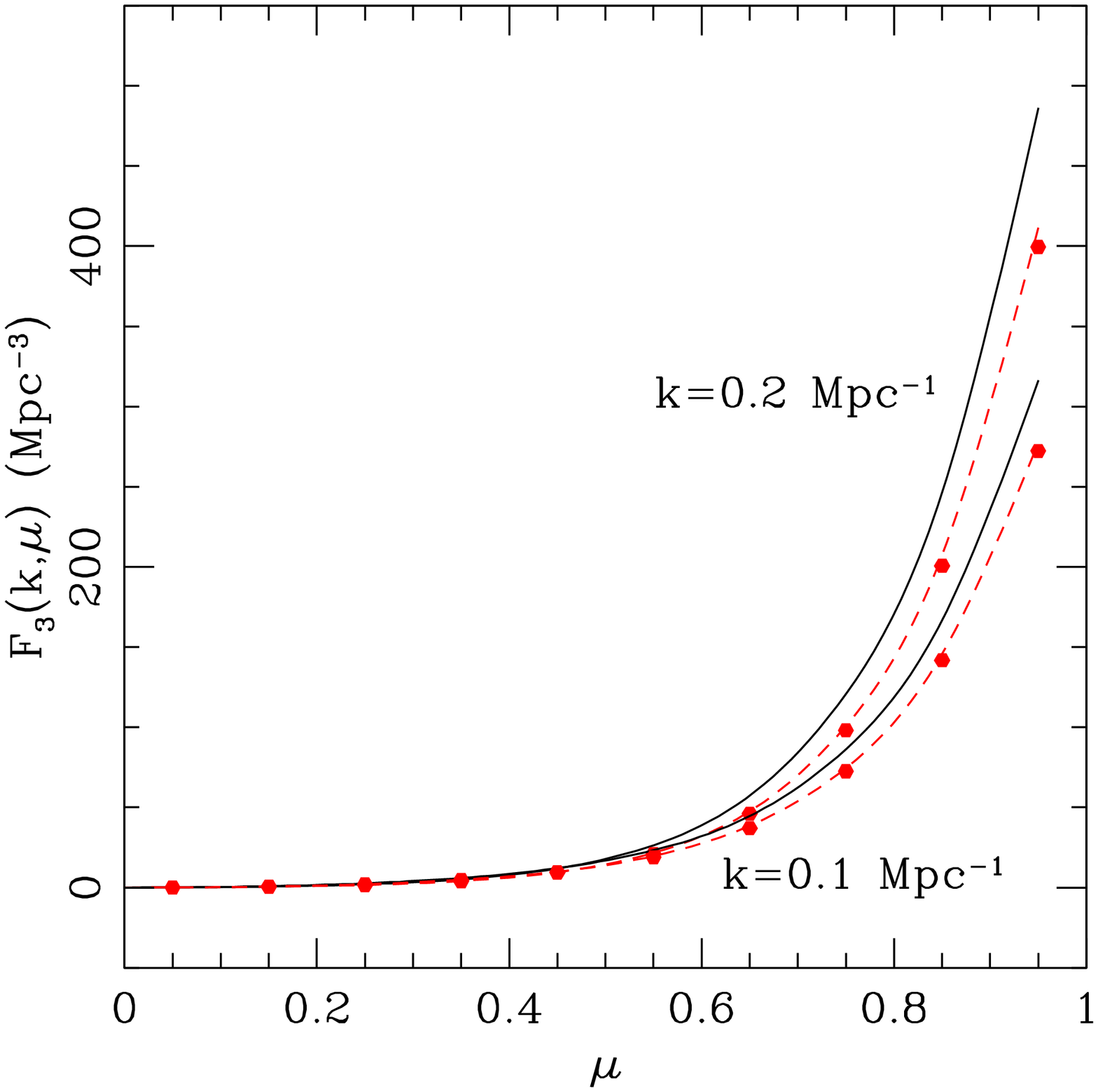}} 
\end{center}
\vspace*{1.0cm}
\caption{Higher-order correction in the redshift-space power spectrum, $F(k,\mu)$, at $z=0.5$. Dividing the function $F$ into three pieces [see Eq.~(\ref{eq:estimatedFn}) and Appendix \ref{app:explicit_form_B_F_T}], the measured $\bar F_n$ are plotted in black solid curves for the fiducial cosmology. The measured $F_n$ is also shown in red dotted points for another cosmological model with $h=0.72$  at the specific scales of $k=0.1$ and $0.2\ompc$. The red dashed curves are the predictions based on the measured $\bar F_n$ in fiducial model using the scaling relation at Eq.~(\ref{eq:estimatedFn}). }
\label{fig:Fkmu}
\end{figure*}

In similar manner to the power spectra $P_{XY}$, we can proceed to a high-precision modeling of the higher-order terms of the redshift-space power spectrum, i.e., $A$, $B$, $T$, and $F$. But, PT prediction for $T$ terms needs a rather higher-order calculation involving the multi-dimensional integrals. Further, in our previous paper
~\cite{Zheng16a}, we directly calibrated each term with $N$-body simulations, and found that the measured $T$ term is different from the PT prediction. For these reasons, we here stick to a $N$-body-based modeling to the higher-order corrections, and adopting the scaling ansatz of the growth factor dependence, the predictions will be made in general cosmological models close to the fiducial $\Lambda$CDM model.

To begin with, let us consider the $A$ term. From the explicit form,  
the $A$ term is divided into six pieces. Here, we specifically write down the expressions in fiducial cosmological model: 
\begin{eqnarray}
  \bar A(k,\mu)
  &=& j_1\,\int d^3\bfx \,\,e^{i\bfk\cdot\bfx}\,\,\langle A_1A_2A_3\rangle_c\nonumber\\
  &=& \sum_{n=1}^{6} \bar {\cal A}_n
\label{eq:A_term}
\end{eqnarray}
Note again that the barred quantities indicate those computed/measured in fiducial cosmological model. The explicit form of $\bar {\cal A}_n$ is given below:
\ba
\bar {\cal A}_1&=& 2j_1\,\int d^3\bfx \,\,e^{i\bfk\cdot\bfx}\,\,\langle u_z(\bfr) \delta(\bfr) \delta(\bfr')\rangle_c, 
\label{eq:A_1}\\
\bar {\cal A}_2  &=& j_1\,\int d^3\bfx \,\,e^{i\bfk\cdot\bfx}\,\,\langle u_z(\bfr)\delta(\bfr)\,\nabla_zu_z(\bfr')\rangle_c, 
\label{eq:A_2}\\
\bar {\cal A}_3  &=& j_1\,\int d^3\bfx \,\,e^{i\bfk\cdot\bfx}\,\,\langle u_z(\bfr)\,\nabla_zu_z(\bfr)\delta(\bfr')\rangle_c, 
\label{eq:A_3}\\
\bar {\cal A}_4 &=& 2j_1\,\int d^3\bfx \,\,e^{i\bfk\cdot\bfx}\,\,\langle u_z(\bfr)\,\nabla_zu_z(\bfr)\,\nabla_zu_z(\bfr')\rangle_c,
\label{eq:A_4}\\
\bar {\cal A}_5  &=& j_1\,\int d^3\bfx \,\,e^{i\bfk\cdot\bfx}\,\,\langle -\delta(\bfr)u_z(\bfr')\,\nabla_zu_z(\bfr')\rangle_c, 
\label{eq:A_5}\\
\bar {\cal A}_6  &=& j_1\,\int d^3\bfx \,\,e^{i\bfk\cdot\bfx}\,\,\langle -\,\nabla_zu_z(\bfr)u_z(\bfr')\delta(\bfr')\rangle_c.
\label{eq:A_6}
\ea
These terms are measured from $N$-body simulations according to Ref.~\cite{Zheng16a}. To apply the measured results to the prediction in other cosmological models, we assume the scaling ansatz, as similarly adopted in the prediction of power spectrum $P_{\rm XY}$. That is, assuming that the scale-dependence of each term is insensitive to the cosmology, the prediction of each term is made by simply rescaling the measured results. The proposition made here is that the time-dependence of each term is approximately determined by the leading-order growth factor dependence of $u_z$ and $\delta$. Then, $A$ term in general cosmological model is expressed as
\ba\label{eq:estimatedAn}
A(k,\mu)
&=& \sum_{n=1}^{6} {\cal A}_n \\
&=&\left(G_\delta/\bar G_\delta\right)^2\left(G_\Theta/\bar G_\Theta\right)
\bar{\cal A}_1
+\left(G_\delta/\bar G_\delta\right)\left(G_\Theta/\bar G_\Theta\right)^2
\bar{\cal A}_2\\
&+&\left(G_\delta/\bar G_\delta\right)\left(G_\Theta/\bar G_\Theta\right)^2
\bar{\cal A}_3
+\left(G_\Theta/\bar G_\Theta\right)^3
\bar{\cal A}_4\\
&+&\left(G_\delta/\bar G_\delta\right)\left(G_\Theta/\bar G_\Theta\right)^2
\bar{\cal A}_5
+\left(G_\delta/\bar G_\delta\right)\left(G_\Theta/\bar G_\Theta\right)^2
\bar{\cal A}_6
\ea
In Fig.~\ref{fig:Akmu}, 
using the measured $\bar{\cal A}_n$ in fiducial model (black solid), the predictions are made in different cosmological model with $h=0.72$, depicted as red dashed lines. These are compared with the direct measurement results, shown as red filled symbols. In each panel of Fig.~\ref{fig:Akmu}, the results are plotted as function of $\mu$ for specific wavenumbers $k=0.1\ompc$ and $k=0.2\ompc$, and we find that the prediction and measurements reasonably agree well with each other. Since the predicted values of ${\cal A}_n$ are found to be sufficiently accurate even at $k=0.2\ompc$, we do not consider any systematics and uncertainties arising from the higher-order growth function dependence.

\begin{figure*}
\begin{center}
\resizebox{2.0in}{!}{\includegraphics{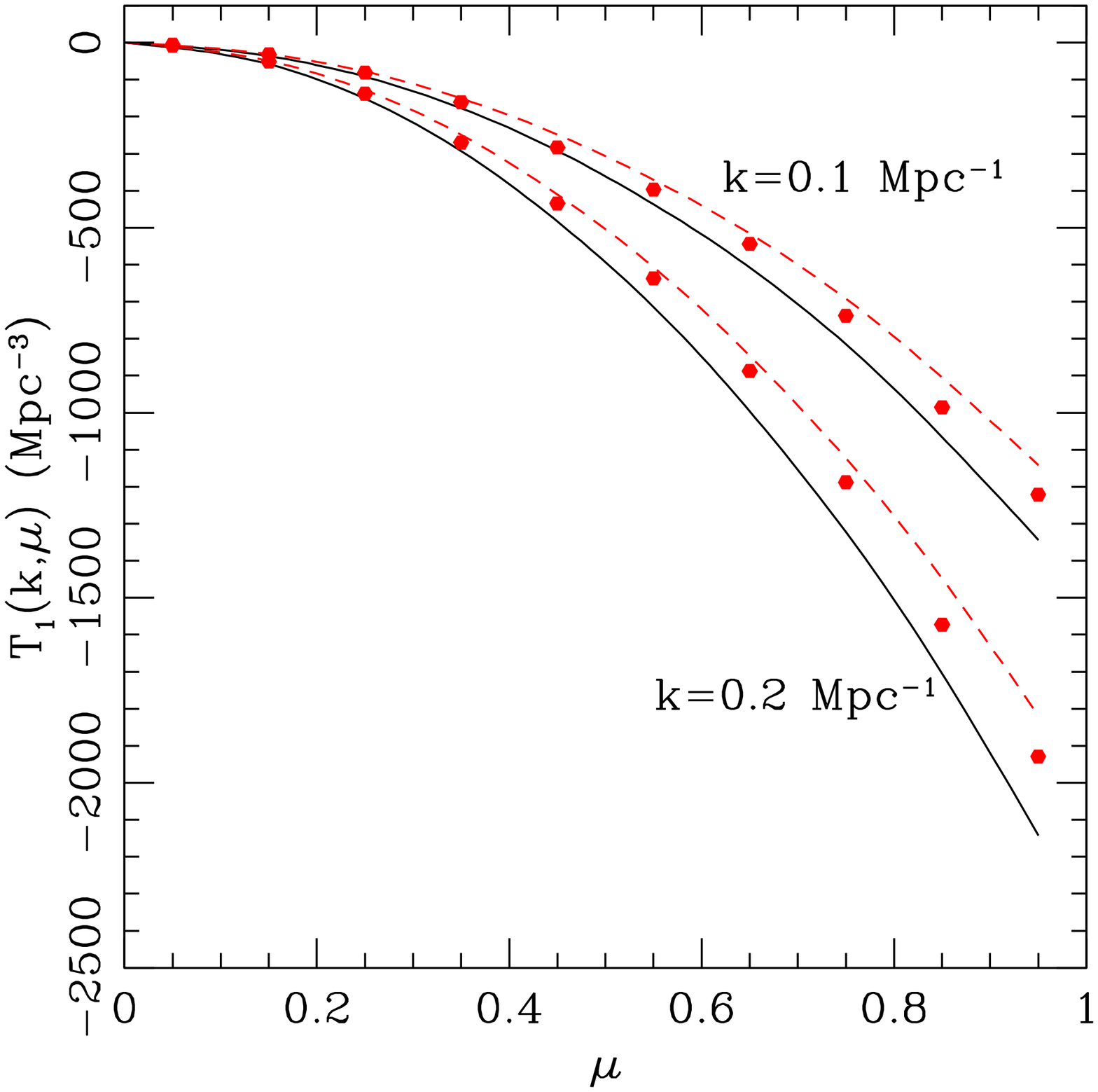}}\hfill
\resizebox{2.0in}{!}{\includegraphics{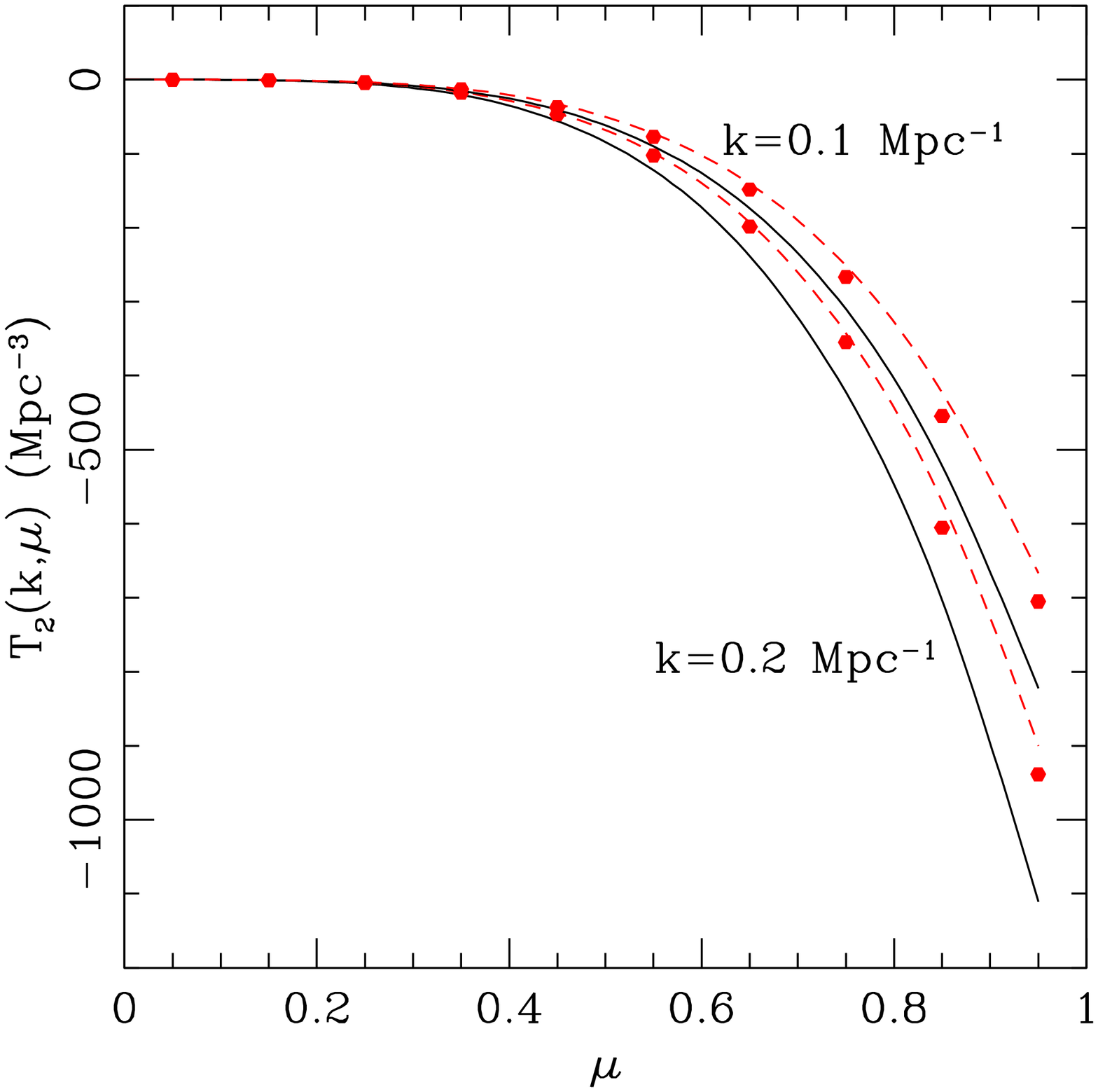}} \hfill
\resizebox{2.0in}{!}{\includegraphics{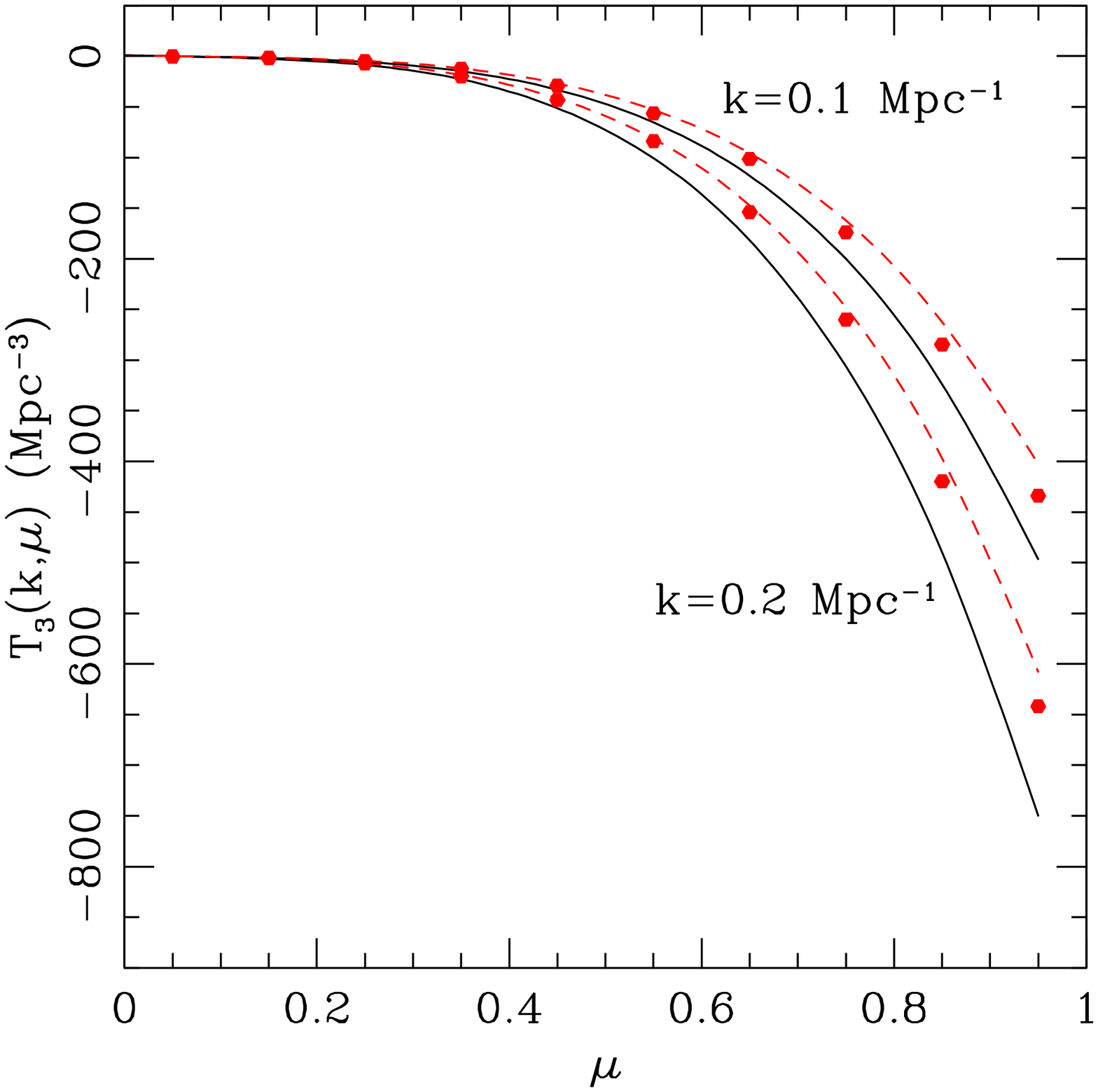}} \\
\resizebox{2.0in}{!}{\includegraphics{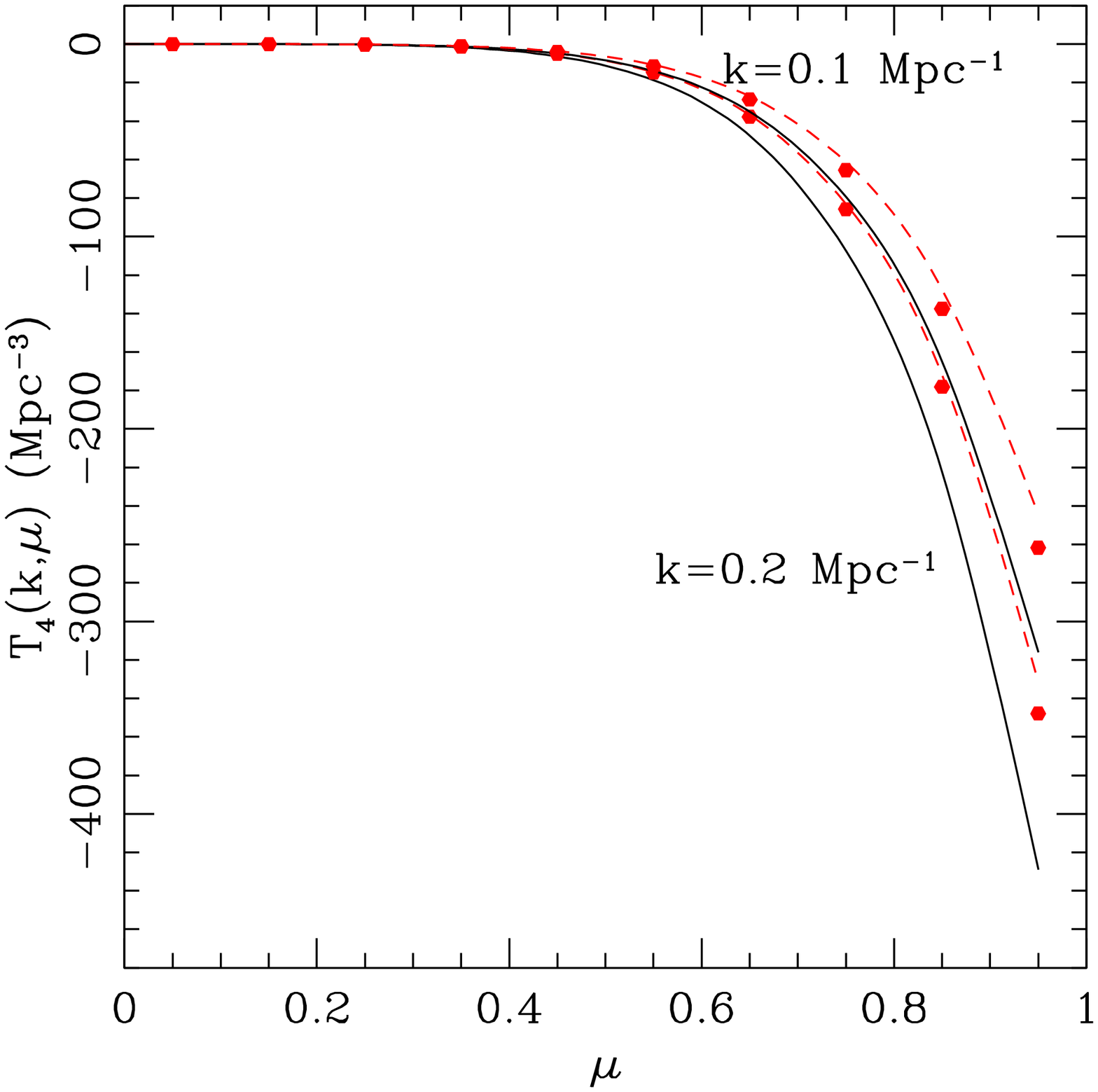}}\hfill
\resizebox{2.0in}{!}{\includegraphics{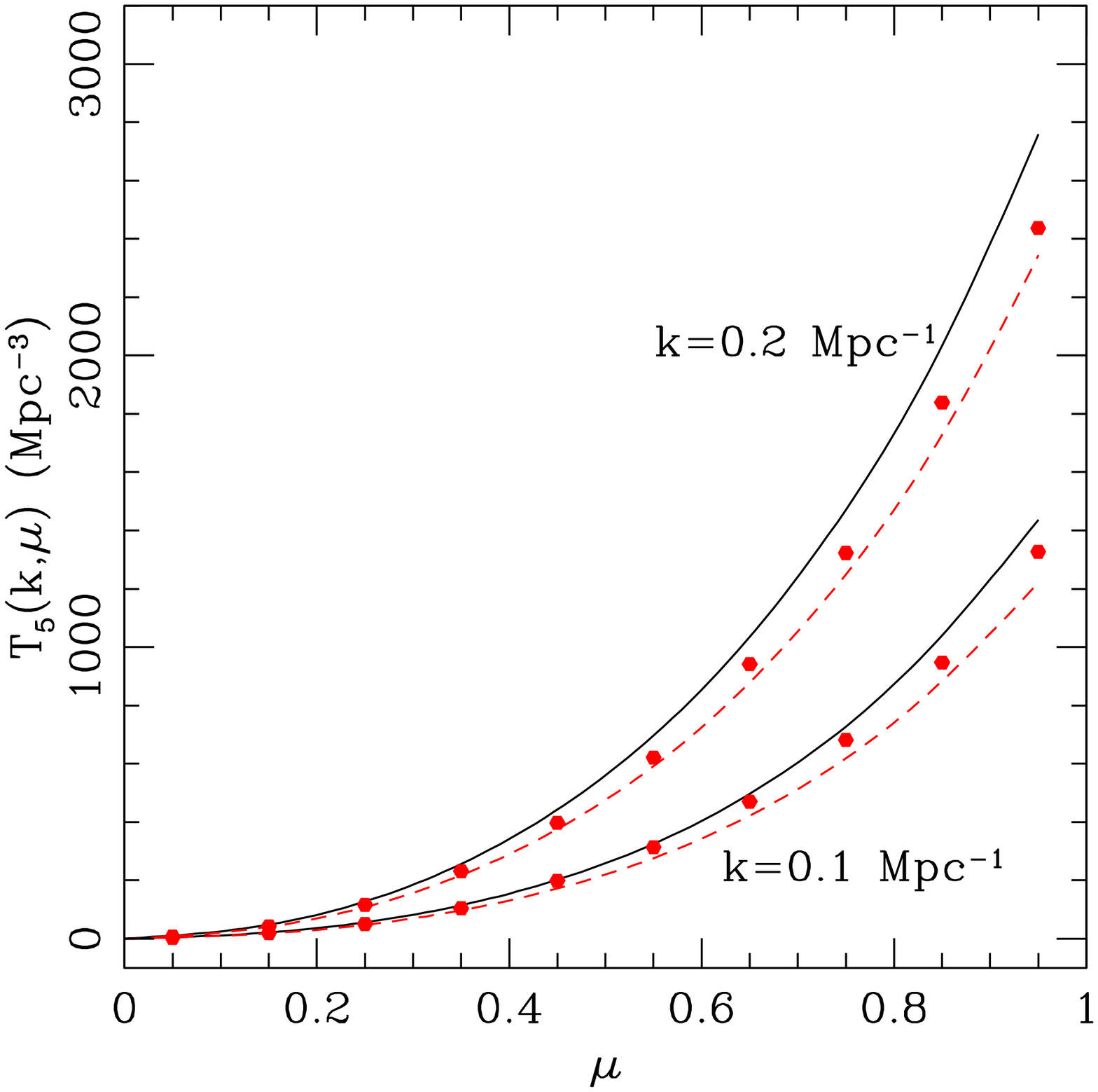}} \hfill
\resizebox{2.0in}{!}{\includegraphics{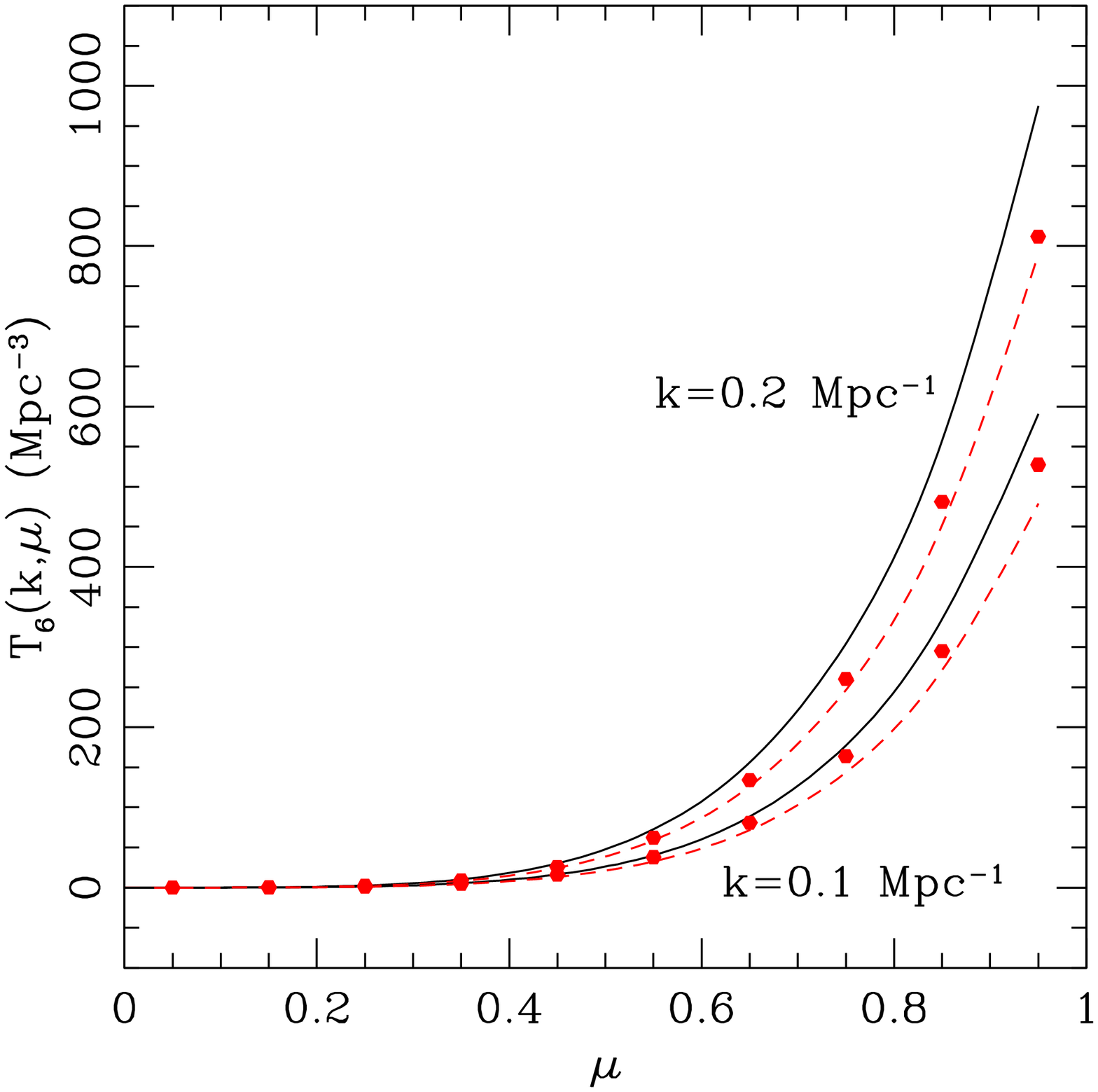}} \\
\resizebox{2.0in}{!}{\includegraphics{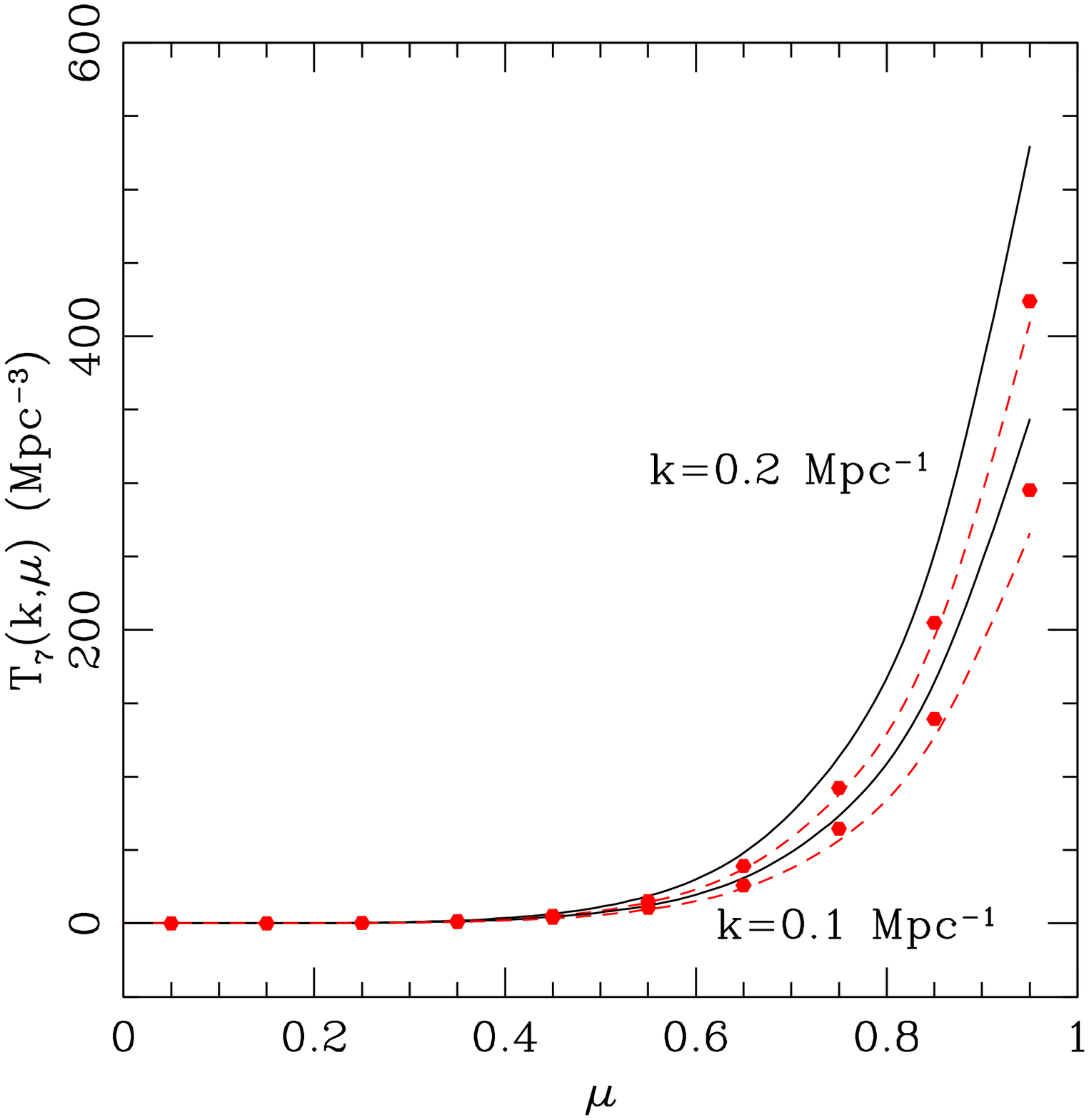}}\hfill
\end{center}
\vspace*{1.0cm}
\caption{Higher-order correction in the redshift-space power spectrum, $T(k,\mu)$, at $z=0.5$. Dividing the function $T$ into seven pieces [see Eq.~(\ref{eq:estimatedTn}) and Appendix \ref{app:explicit_form_B_F_T}], the measured $\bar F_n$ are plotted in black solid curves for the fiducial cosmology. The measured $T_n$ is also shown in red dotted points for another cosmological model with $h=0.72$  at the specific scales of $k=0.1$ and $0.2\ompc$. The red dashed curves are the predictions based on the measured $\bar T_n$ in fiducial model using the scaling relation at Eq.~(\ref{eq:estimatedTn}). }
\label{fig:Tkmu}
\end{figure*}

We then apply the same strategy to other higher-order corrections, $B$, $F$ and $T$. Dividing these corrections into several pieces, the scaling ansatz leads to the following predictions:
\ba
B(k,\mu) &=& \sum_{n=1}^{4} {\cal B}_n 
\label{eq:estimatedBn}
\\
&=& \left(G_\delta/\bar G_\delta\right)^2\left(G_\Theta/\bar G_\Theta\right)^2
\bar{\cal B}_1
+\left(G_\delta/\bar G_\delta\right)\left(G_\Theta/\bar G_\Theta\right)^3
\bar{\cal B}_2
\nn 
\\
&+&\left(G_\delta/\bar G_\delta\right)\left(G_\Theta/\bar G_\Theta\right)^3
\bar{\cal B}_3
+\left(G_\Theta/\bar G_\Theta\right)^4
\bar{\cal B}_4
\nn
\\
F(k,\mu) &=& \sum_{n=1}^{3} {\cal F}_n 
\label{eq:estimatedFn}
\\
&=& \left(G_\delta/\bar G_\delta\right)^2\left(G_\Theta/\bar G_\Theta\right)^2
\bar{\cal F}_1
+\left(G_\delta/\bar G_\delta\right)\left(G_\Theta/\bar G_\Theta\right)^3
\bar{\cal F}_2
+\left(G_\Theta/\bar G_\Theta\right)^4
\bar{\cal F}_3
\nn
\\
T(k,\mu)&=& \sum_{n=1}^{7} {\cal T}_n 
\label{eq:estimatedTn}
\\
&=& \left(G_\delta/\bar G_\delta\right)^2\left(G_\Theta/\bar G_\Theta\right)^2
\bar{\cal T}_1
+\left(G_\delta/\bar G_\delta\right)\left(G_\Theta/\bar G_\Theta\right)^3
\bar{\cal T}_2
+\left(G_\delta/\bar G_\delta\right)\left(G_\Theta/\bar G_\Theta\right)^3
\bar{\cal T}_3
\nn
\\
&+&\left(G_\Theta/\bar G_\Theta\right)^4
\bar{\cal T}_4
+\left(G_\delta/\bar G_\delta\right)^2\left(G_\Theta/\bar G_\Theta\right)^2
\bar{\cal T}_5
+\left(G_\delta/\bar G_\delta\right)\left(G_\Theta/\bar G_\Theta\right)^3
\bar{\cal T}_6
+\left(G_\Theta/\bar G_\Theta\right)^4
\bar{\cal T}_7 
\nn
\ea
Here, the quantities $\bar{\cal B}_n$, $\bar{\cal F}_n$, and 
$\bar{\cal T}_n$ are measured in the fiducial cosmological model. Definition and explicit form of each quantity is presented in Appendix \ref{app:explicit_form_B_F_T}.
Figs.~\ref{fig:Bkmu}, \ref{fig:Fkmu} and \ref{fig:Tkmu} respectively show the quantities ${\cal B}_n$, ${\cal F}_n$, and ${\cal T}_n$, plotted as function of $\mu$. As similarly shown in Fig.~\ref{fig:Akmu}, for specific wavenumbers $k=0.1\ompc$ and $0.2\ompc$, the measured results in fiducial model are depicted as black solid lines, while the the predictions based on the scaling ansatz are shown in red dashed lines, which are compared with direct measurements (red filled circles) in the cosmological model with a slightly different value of $h$. The results show that the simple scaling ansatz also works well for all higher-order corrections, and suggests that the approach examined here can be used as a high-precision template, at least at $k\lesssim0.2\ompc$.

\section{Testing power spectrum template: accurate estimation of growth functions}


\begin{figure}
\begin{center}
\resizebox{3.2in}{!}{\includegraphics{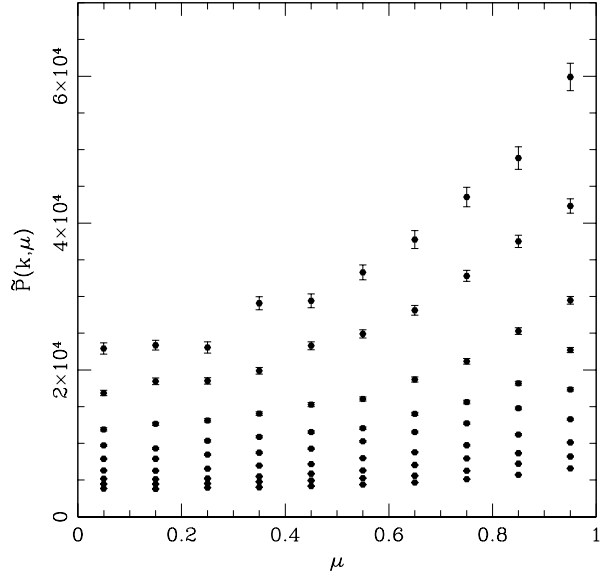}}
\end{center}
\vspace*{1.5cm}
\caption{Redshift-space power spectrum at $z=0.5$, measured from $N$-body simulations. The mean values of the power spectrum, $\tilde{P}_{\rm ob}(k,\mu)$, are estimated using $100$ realizations of the simulation data, and results are plotted as function of directional cosine $\mu$ at the wavenumbers $k=0.055$, $0.075$, $0.095$, $0.115$, $0.135$, $0.155$, $0.175$, $0.195$, and $0.215\hompc$ (from top to bottom), together with their $1\sigma$ error.}
\label{fig:measuredPkmu}
\end{figure}

In this section, we test our hybrid theoretical template, and applying it to the simulation data of power spectrum, we demonstrate that the best-fitted values of the growth functions consistently reproduce those assumed in the simulations.

\subsection{Parameter estimation}
\label{subsec:parameter_estimation}

We first measure the power spectrum in 2D space from the simulations for each cosmological model, which is later fitted by the theoretical template. The output data of dark matter particles at $z=0.5$ 
are taken from the $100$ realizations of the simulations as we described in Sec.~\ref{subsec:high-precison_PXY}. The measurement is then performed in grid space from the grid-assigned density field using the fast Fourier transform, and the resultant data are stored in $k$ and $\mu$ bins at $0.01\hompc <k< 0.3\hompc$ and $0<\mu<1$ with the bin size of $\Delta k=0.01\hompc$ and $\Delta \mu =0.1$.  In Fig.~\ref{fig:measuredPkmu},  the measured result of the power spectrum is presented in fiducial cosmological model, at the specific wavenumber bins of $k=0.055$, $0.075$, $0.095$, $0.115$, $0.135$, $0.155$, $0.175$, $0.195$, and $0.215$ $\hompc$ (from top to bottom). 
Provided the power spectrum data in each cosmological model, we next compare it to the theoretical template given at Eq.~(\ref{eq:Pkred_final}), and the growth functions for density and velocity fields, $G_\delta$ and $G_\Theta$, are estimated, marginalizing over the nuisance parameter $\sigma_z$ that characterizes the strength of FoG damping, for which we assume the Gaussian form [see Eq.~(\ref{eq:RSD_model_extended})]. Thus, the number of parameters to be determined is three. Since the power spectrum template involves several correction terms in Eq.~(\ref{eq:Pkred_final}), we will examine different combinations of $A$, $B$, $F$ and $T$ terms to see which combination gives the best estimate for the growth functions $G_\delta$ and $G_\Theta$. The combinations we examined below include, $A+B$ (theory), $A+B$, $A+B+T$, and $A+B+F+T$. The analysis of parameter estimation is performed with Markov chain Monte Carlo technique based on the $\chi^2$ given by
\ba\label{eq:chisq}
\chi^2=\sum^{i_{\rm max}}_{i=i_{\rm min}}\sum^{10}_{p=1} \sum^{10}_{q=1}  
[\tilde{P}_{\rm obs}(k_i,\mu_p)-\tilde{P}_{\rm model}(k_i,\mu_p)]
{\rm Cov}^{-1}_{pq}(k_i)
[\tilde{P}_{\rm obs}(k_i,\mu_q)-\tilde{P}_{\rm model}(k_i,\mu_q)]\,,
\ea
where the quantities $\tilde{P}_{\rm obs}$ and $\tilde{P}_{\rm model}$ are the measured and template power spectra, respectively . Setting the minimum wavenumber $k_{\rm min}$ to $0.01\hompc$, we will examine the parameter estimation varying $k_{\rm max}$. Here, the error covariance of the measured power spectrum, ${\rm Cov}$, is assumed to be described by the Gaussian covariance with the non-vanishing diagonal element given by $\sigma[\tilde{P}_{\rm obs}(k,\mu)]=\tilde{P}_{\rm obs}(k,\mu)\sqrt{2/N(k,\mu)}$, where $N(k,\mu)$ is the number of Fourier modes in each $(k,\mu)$-bin. Then, the inverse covariance, ${\rm Cov}^{-1}$, becomes
\ba
{\rm Cov}^{-1}_{pp}(k_i)=\frac{1}{\sigma[\tilde{P}_{\rm ob}(k_i,\mu_p)]^2}\,.
\ea

\begin{figure*}
\begin{center}
\resizebox{3.in}{!}{\includegraphics{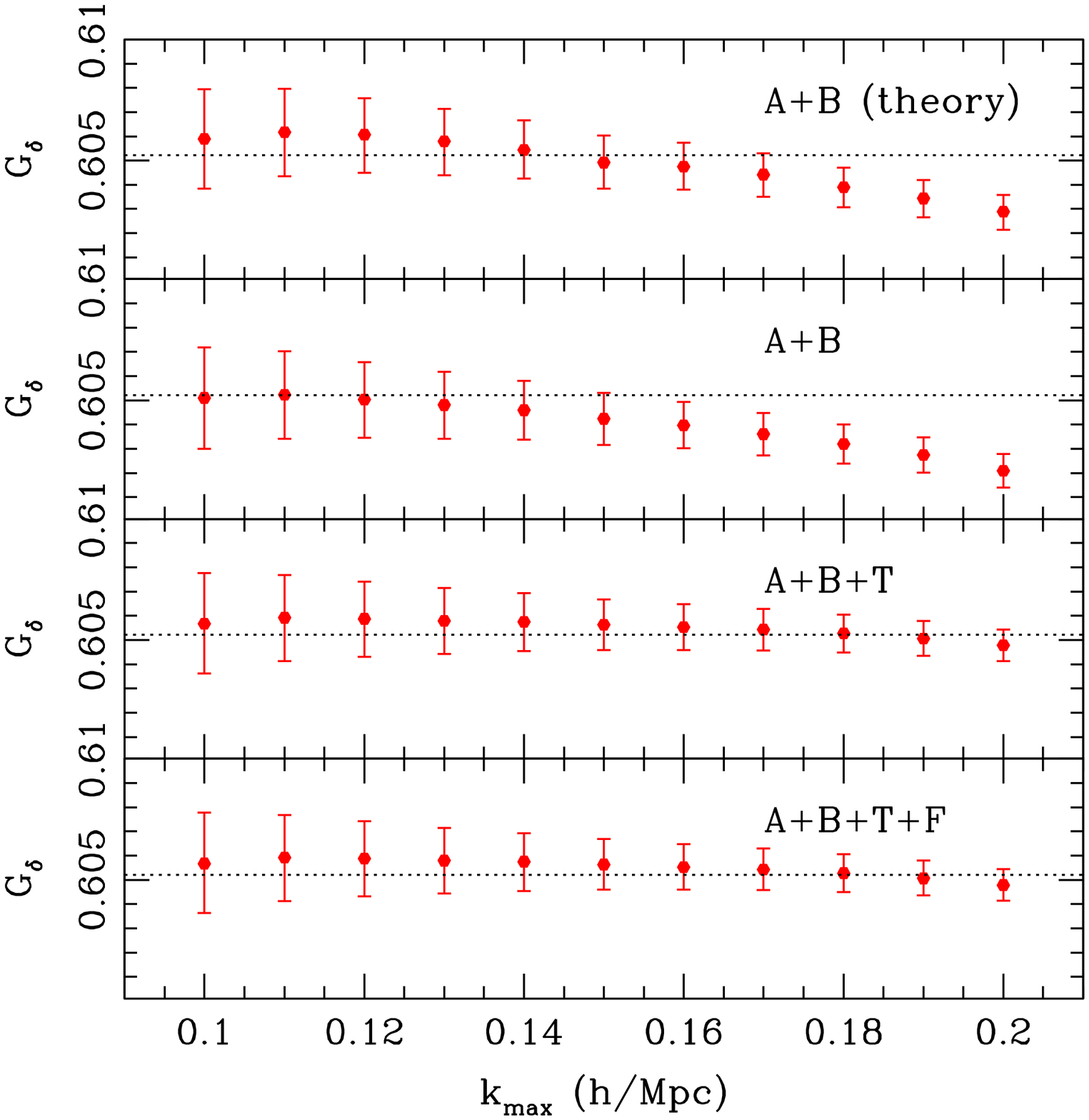}}
\resizebox{3.in}{!}{\includegraphics{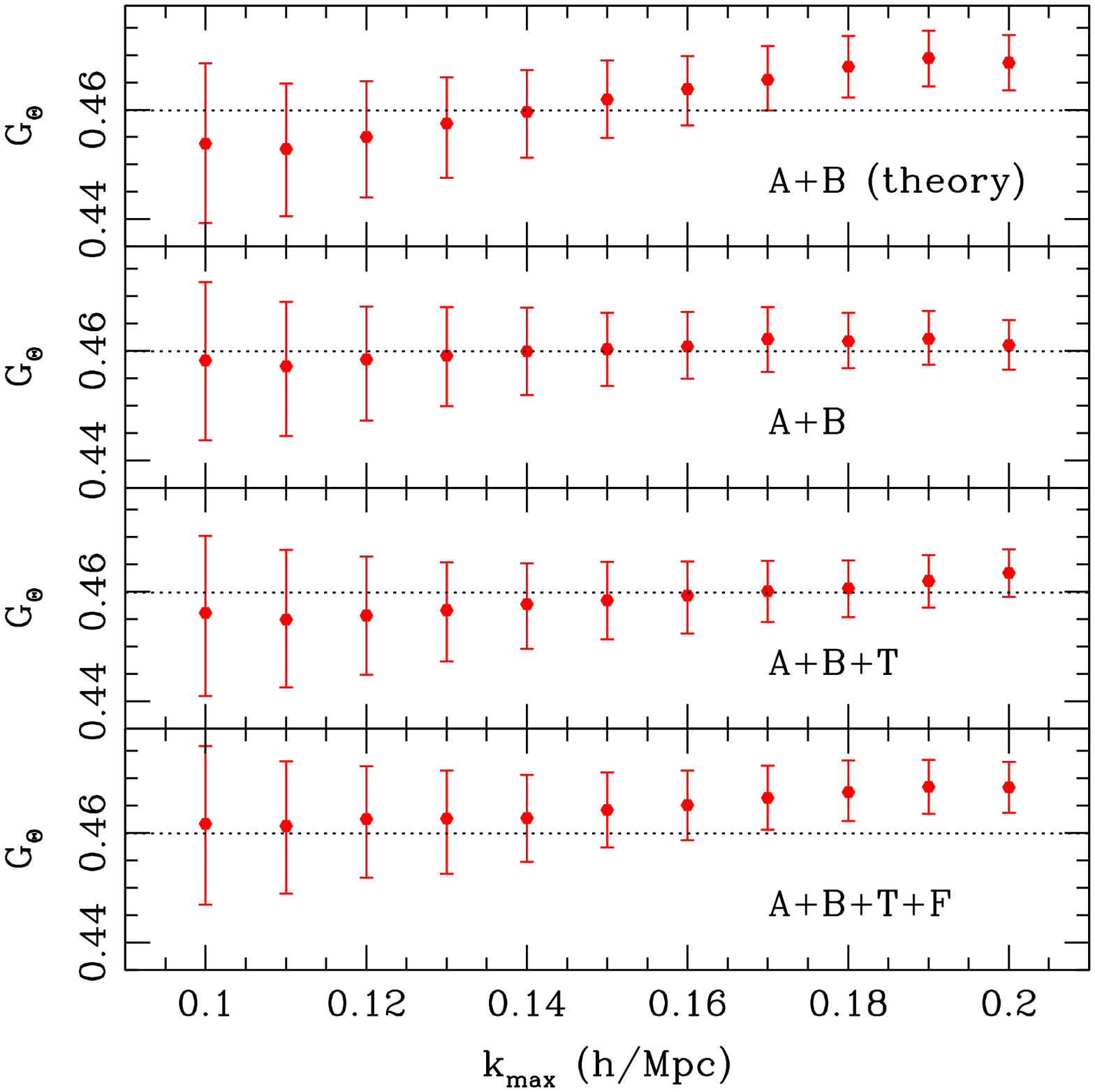}}
\end{center}
\vspace*{1.5cm}
\caption{Best-fit values of $G_\delta$ (left) and $G_\Theta$ (right) as function of $k_{\rm max}$ at $z=0.5$. The errorbars indicate the $1\sigma$ statistical error among $100$ realizations. The results are obtained from theoretical templates with various combinations of correction terms. From top to bottom, the template include the corrections $A+B$ (theory),  $A+B$, $A+B+T$, and $A+B+T+F$. Note that these correction terms were measured from $N$-body simulations in fiducial  cosmology except for top panel, where PT treatment is used to evaluate $A+B$ terms. }
\label{fig:measuredG}
\end{figure*}

\begin{figure*}
\begin{center}
\resizebox{3.in}{!}{\includegraphics{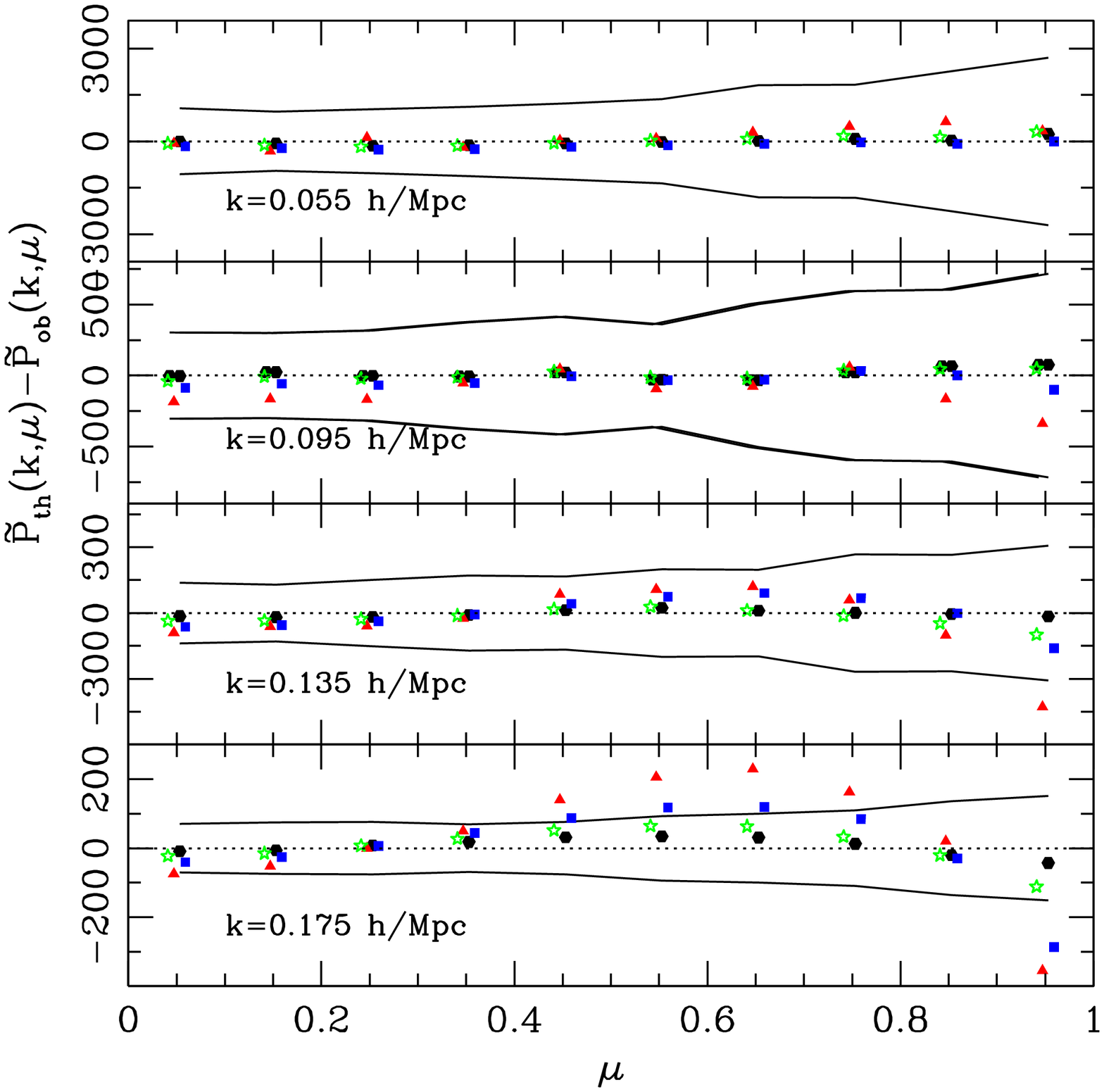}}
\resizebox{3.in}{!}{\includegraphics{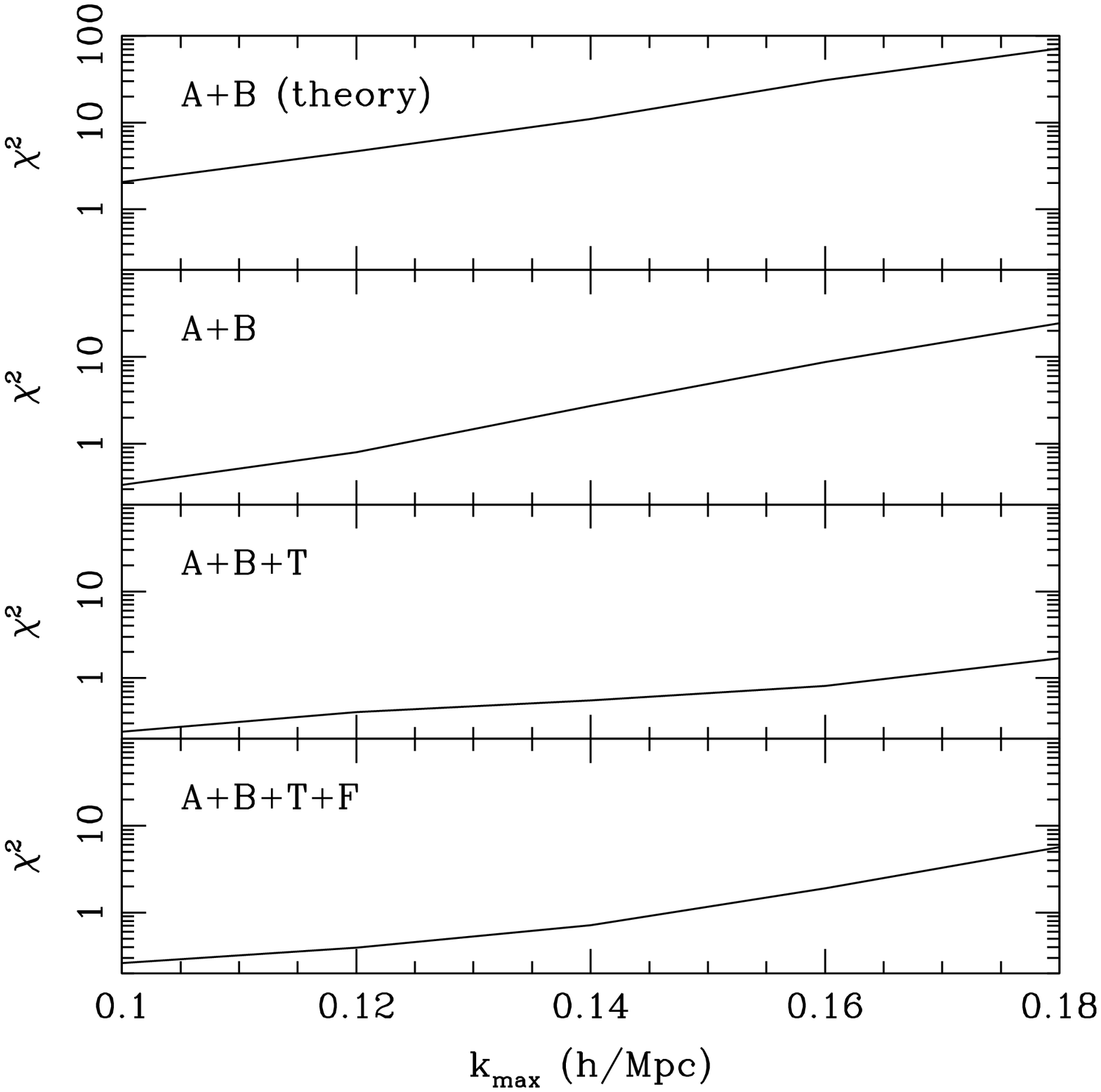}}
\end{center}
\vspace*{1.5cm}
\caption{{\it Left}: Residuals of the best-fit power spectrum, i.e., $\tilde{P}_{\rm model}(k_i,\mu_p)-\tilde{P}_{\rm obs}(k_i,\mu_p)$, plotted as function of $\mu$ at $k=0.055$, $0.095$, $0.135$, $0.175\hompc$. For the templates with $A+B$ (theory), $A+B$, $A+B+T$ and $A+B+T+F$, the results are shown in red-triangles, blue squares, black circles, and green star dots, respectively. The solid curves represent the $1\sigma$ errors estimated from $100$ simulations. {\it Right}: The $\chi^2$ values of the best-fit power spectra for various combinations of the correction terms shown in left panels. From top to bottom, the results for the templates with $A+B$ (theory), $A+B$, $A+B+T$ and $A+B+T+F$ are respectively plotted against $k_{\max}$. }
\label{fig:growthABTF}
\end{figure*}

\subsection{Results}

Let us first see if our modeling of RSD really works well to describe the power spectrum without any systematics and uncertainty. To do this, we examine the parameter estimation in the fiducial cosmological model, and instead of employing the PT calculations, we stick to the $N$-body data to compute the theoretical template not only for the correction terms (i.e., $A$, $B$, $T$ and $F$ terms) but also for the power spectra $\bar P_{\delta\delta}$, $\bar P_{\delta\Theta}$ and $\bar P_{\Theta\Theta}$.

Fig.~\ref{fig:measuredG} shows the best-fit values and their $1\sigma$ marginalized error for the growth functions $G_\delta$ (left) and $G_\Theta$ (right), plotted against $k_{\rm max}$. The results are summarized for the theoretical templates with different combinations of the correction terms: $A+B$, $A+B+T$ and $A+B+F+T$ (from the second top to bottom). For reference, the top panels similarly show the cases when the template includes the corrections $A+B$, but different from the second top panels, $A+B$ terms are here obtained from the PT calculations. Then, the best performance of the parameter estimation comes from the template with $A+B+T$ combination, not with the full $A+B+F+T$ combination. The reason for this may be ascribed to the multi-streaming motion of dark matter particles, which approximately cancels the higher-order correlation $F$ \cite{Zheng16c}. The template with $A+B$ gives a poorer estimation for the growth function $G_\delta$, and employing the PT calculations, the discrepancy becomes prominent even for $G_\theta$, partly due to the systematics in the PT prediction.

To see more quantitatively whether the best-fit results of the theoretical template consistently reproduce the measured power spectrum or not,  we evaluate the differences between the best-fit and measured results, i.e., $\tilde{P}_{\rm model}-\tilde{P}_{\rm obs}$, and plot the residuals in left panels of Fig.~\ref{fig:growthABTF}. The red triangle, blue square, black circle and green star dots represent the results based on the template including $A+B$ (theory), $A+B$, $A+B+F$ and $A+B+F+T$, respectively. At small $\mu$, most of the results are within the statistical error depicted as solid black curves, however, a large discrepancy is manifest at $\mu\to1$, where the non-perturbative effect of RSD becomes significant. As a result, the template including $A+B+T$ shows the best performance. This is indeed manifest if we evaluate the best-fit $\chi^2$, shown in right panels of Fig.~\ref{fig:growthABTF}.  Hence, we conclude that with the template including $A+B+T$ terms, unbiased estimation of the growth functions $G_\delta$ and $G_\Theta$ is possible at $k_{\rm max}\lesssim0.18\hompc$, below which the best-fit result consistently reproduces the measured power spectrum.

\begin{figure}
\begin{center}
\resizebox{3.2in}{!}{\includegraphics{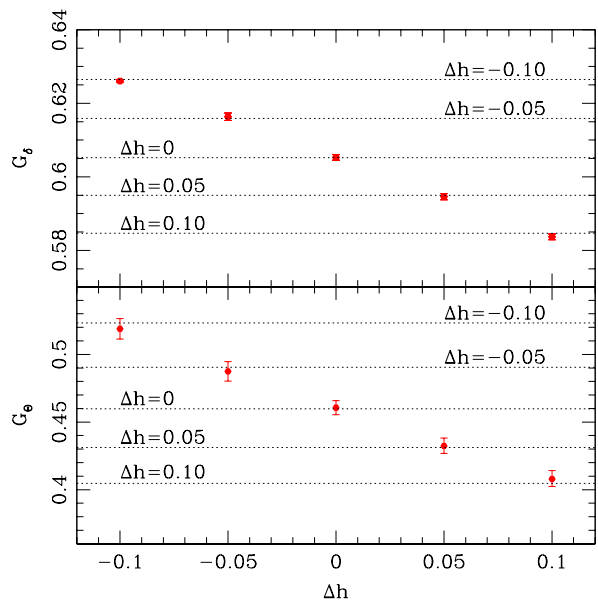}}
\end{center}
\vspace*{1.5cm}
\caption{Best-fit values of the growth functions, $G_\delta$ (top) and $G_\Theta$ (bottom), for cosmological models with different value of $h$. Here, based on the template with the correction terms $A+B+T$, the growth functions are determined by fitting the template to the measured power spectrum at $k\lesssim k_{\rm max} = 0.18\hompc$. The resultant best-fit values are shown as function of the difference of $h$ from the fiducial cosmology, $\Delta h$. For reference, the dotted lines represent the fiducial values of growth functions for each cosmological model.}
\label{fig:allcosmology}
\end{figure}

Let us then demonstrate that using the scaling relation, our calibrated template in the fiducial cosmology can also work well in other cosmological models. Fig.~\ref{fig:allcosmology} shows the results of the parameter estimation in cosmological models with different value of $h$. Here, the template power spectrum $\tilde{P}_{\rm ob}(k_i,\mu_p)$ including the $A+B+T$ corrections is computed in the fiducial cosmology of $h=0.67$, and using the scaling ansatz, the growth functions are determined by fitting the template to the measured power spectrum at $k\lesssim k_{\rm max}=0.18 \hompc$. The best-fit value of the growth functions recovers the one in the cosmological model, and the accuracy of the parameter estimation reaches at 1\% level. Although the present study restricts the analysis in the $\Lambda$CDM models just varying $h$, the power spectrum template as well as our methodology can be applied to a wide class of cosmological models as long as the broadband power spectrum shape remains the same one as in the $\Lambda$CDM model, and the scaling relation in previous section holds. Thus, Fig.~\ref{fig:allcosmology} implies that the accuracy to estimate the growth functions is expected to hold in general dark energy and/or modified gravity models, in which the cosmic expansion and (scale-independent) growth of structure are different from $\Lambda$CDM predictions. 

\section{Conclusion}

In this paper, with the aid of both the PT calculations and $N$-body simulations, we have presented a hybrid model of redshift-space power spectrum, which enables us to estimate the growth functions with 1\% accuracy. Adopting the treatment by Ref.~\cite{Taruya10,Taruya13}, we first presented the model in which the non-perturbative effect of RSD is separated out, and the rest of the contributions can be computed perturbatively. To make any systematic uncertainty or small flaw under control, we consider the approach by Ref.~\cite{Zheng16a}, and each of the contributions in the power spectrum template has been calibrated or measured with $N$-body simulations in fiducial cosmological model. Using the scaling relation for the dependence of the growth functions, the calibrated template is then applied to other cosmological models in which the broadband power spectrum is basically the same one as in the $\Lambda$CDM model, relevant to general dark energy models and/or a class of modified gravity models that can change both the cosmic expansion and the growth of structure in a scale-independent manner. 

We demonstrated that fitting our hybrid template to the power spectrum data in $N$-body simulations, the unbiased estimation of the growth function is possible. In particular, with the template including the corrections $A+B+T$, the best-fit results of the growth functions are shown to reach at 1\% accuracy at $k_{\rm max}\lesssim 0.18\hompc$. Although all the analysis in the paper has been made at $z=0.5$, we expect that the performance of our template basically remains the same or becomes even better at higher redshifts. With the upcoming galaxy surveys like DESI or Euclid, a severe cosmological test of gravity is thus made possible with our hybrid template.

Toward practical application, however, one crucial step is to properly incorporate the effect of galaxy bias into the power spectrum template. To tackle this issue, our approach has to be tested against the halo or mock galaxy catalogs based on an appropriate prescription for halo/galaxy bias \cite{Zheng18a}. Another important generalization is to seek a flexible template in which the broadband shape of the power spectrum is allowed to vary. To do this, one needs to exploit the fast PT calculation as well as to find a more cleaver way to calibrate the power spectrum with $N$-body simulation. We will hopefully report our progress near future.

\section*{Acknowledgements}

To complete this work, discussions during the workshop, YITP-T-17-03, held at Yukawa Institute for Theoretical Physics at Kyoto University were very useful. 
This work is supported in part by MEXT/JSPS KAKENHI Grant Number JP15H05899 and JP16H03977 (AT). The work of running simulation was supported by the National Institute of Supercomputing and Network/Korea Institute of Science and Technology Information with supercomputing resources including technical support (KSC-2017-C1-0006). Numerical calculations were performed by using a high performance computing cluster in the Korea Astronomy and Space Science Institute.

\appendix
\section{Explicit functional form of ${\cal B}_n$, ${\cal F}_n$, and ${\cal T}_n$}
\label{app:explicit_form_B_F_T}

In this Appendix, we present the explicit form of the scale-dependent coefficients for the higher-order correction terms, $B$, $T$, and $F$, given at Eqs.~(\ref{eq:estimatedBn}), (\ref{eq:estimatedFn}), and (\ref{eq:estimatedTn}). 

Here, we write down the expressions in the fiducial cosmological model. First, the $B$ term is divided into four pieces:
\begin{eqnarray}
  \bar B(k,\mu)&=&  j_1^2\,\int d^3\bfx \,\,e^{i\bfk\cdot\bfx}\,\,\langle A_1A_2\rangle_c\,\langle A_1A_3\rangle_c\nonumber\\
  &=& \sum_{n=1}^{4} \bar {\cal B}_n
\end{eqnarray}
with the terms ${\cal B}_n$ given by
\ba
\bar {\cal B}_1&=&  j_1^2\,\int d^3\bfx \,\,e^{i\bfk\cdot\bfx}\,\,\langle -u_z(\bfr')\delta(\bfr)\rangle_c\,\langle u_z(\bfr)\delta(\bfr')\rangle_c\\
\bar {\cal B}_2  &=&  j_1^2\,\int d^3\bfx \,\,e^{i\bfk\cdot\bfx}\,\,\langle -u_z(\bfr')\delta(\bfr)\rangle_c\,\langle u_z(\bfr)\,\nabla_zu_z(\bfr')\rangle_c\\
\bar {\cal B}_3 &=& j_1^2\,\int d^3\bfx \,\,e^{i\bfk\cdot\bfx}\,\,\langle -u_z(\bfr')\,\nabla_zu_z(\bfr)\rangle_c\,\langle u_z(\bfr)\delta(\bfr')\rangle_c\\
\bar {\cal B}_4 &=& j_1^2\,\int d^3\bfx \,\,e^{i\bfk\cdot\bfx}\,\,\langle -u_z(\bfr')\,\nabla_zu_z(\bfr)\rangle_c\,\langle u_z(\bfr)\,\nabla_zu_z(\bfr')\rangle_c
\ea
The $T$ term is divided into seven pieces:
\begin{eqnarray}
  \bar T(k,\mu)&=&\frac{1}{2} j_1^2\,\int d^3\bfx \,\,e^{i\bfk\cdot\bfx}\,\,\langle A_1^2A_2A_3\rangle_c,\nonumber\\
  &=& \sum_{n=1}^{7} \bar {\cal T}_n\,;
\end{eqnarray}
\ba
\bar {\cal T}_1&=&   j_1^2\,\int d^3\bfx \,\,e^{i\bfk\cdot\bfx}\,\,\langle u_z(\bfr)u_z(\bfr)\delta(\bfr)\delta(\bfr') \rangle_c\\
\bar {\cal T}_2  &=& j_1^2\,\int d^3\bfx \,\,e^{i\bfk\cdot\bfx}\,\,\langle u_z(\bfr)u_z(\bfr)\delta(\bfr)\,\nabla_zu_z(\bfr')  \rangle_c\\
\bar {\cal T}_3  &=&  j_1^2\,\int d^3\bfx \,\,e^{i\bfk\cdot\bfx}\,\,\langle u_z(\bfr)u_z(\bfr)\,\nabla_zu_z(\bfr)\delta(\bfr')  \rangle_c\\
\bar {\cal T}_4 &=& j_1^2\,\int d^3\bfx \,\,e^{i\bfk\cdot\bfx}\,\,\langle u_z(\bfr)u_z(\bfr)\,\nabla_zu_z(\bfr)\,\nabla_zu_z(\bfr')  \rangle_c\\
\bar {\cal T}_5&=&  \frac{1}{2} j_1^2\,\int d^3\bfx \,\,e^{i\bfk\cdot\bfx}\,\,\langle -2u_z(\bfr')u_z(\bfr)\delta(\bfr)\delta(\bfr') \rangle_c\\
\bar {\cal T}_6  &=& j_1^2\,\int d^3\bfx \,\,e^{i\bfk\cdot\bfx}\,\,\langle -2u_z(\bfr')u_z(\bfr)\delta(\bfr)\,\nabla_zu_z(\bfr')  \rangle_c \\
\bar {\cal T}_7 &=& \frac{1}{2} j_1^2\,\int d^3\bfx \,\,e^{i\bfk\cdot\bfx}\,\,\langle -2u_z(\bfr')u_z(\bfr)\,\nabla_zu_z(\bfr)\,\nabla_zu_z(\bfr')  \rangle_c  
\ea
Finally, the $F$ term is divided into three pieces:
\ba
\bar F(k,\mu)&=& -j_1^2\,\int d^3\bfx \,\,e^{i\bfk\cdot\bfx}\,\,\langle u_z u_z'\rangle_c\langle A_2A_3\rangle_c
\nn\\
&=& \sum_{n=1}^{3} \bar {\cal F}_n\,\,;
\ea
\ba
\bar{\cal F}_1&=& -j_1^2\,\int d^3\bfx \,\,e^{i\bfk\cdot\bfx}\,\,\langle u_z u_z'\rangle_c\langle \delta(\bfr)\delta(\bfr')\rangle_c\\
\bar{\cal F}_2&=& -2j_1^2\,\int d^3\bfx \,\,e^{i\bfk\cdot\bfx}\,\,\langle u_z u_z'\rangle_c\langle \delta(\bfr)\,\nabla_zu_z(\bfr')\rangle_c\\
\bar{\cal F}_3&=& -j_1^2\,\int d^3\bfx \,\,e^{i\bfk\cdot\bfx}\,\,\langle u_z u_z'\rangle_c\langle \,\nabla_zu_z(\bfr)\,\nabla_zu_z(\bfr')\rangle_c.
\ea



\bibliographystyle{JHEP}

\providecommand{\href}[2]{#2}\begingroup\raggedright\endgroup

\end{document}